\pdfoutput=1

\documentclass[12pt,aps,onecolumn,preprint,nofootinbib]{revtex4}

\usepackage{times}

\usepackage{amsmath}
\usepackage{graphicx}
\renewcommand{\L}{\mathcal{L}}
\renewcommand{\H}{\mathcal{H}}

\newcommand{\figref}[1]{FIG.~\ref{#1}}
\newcommand{\Input}{I}
\newcommand{\vecx}{\mathbf{x}}
\newcommand{\set}[1]{\{#1\}}
\newcommand{\best}{\mathrm{best}}
\newcommand{\vecy}{\mathbf{y}}
\usepackage{bibunits}

\begin{document}

\title{Automated adaptive inference of coarse-grained dynamical models
  in systems biology}


\author{Bryan C.~Daniels$^{\ast}$}
\affiliation{Center for Complexity and Collective Computation, Wisconsin Institute for Discovery,
University of Wisconsin, Madison, WI 53715, USA}

\author{Ilya Nemenman$^{\ast}$}
\affiliation{Departments of Physics and Biology\\
Emory University, Atlanta, GA 30322, USA\\}

\begin{abstract}
  
  Cellular regulatory dynamics is driven by large and intricate
  networks of interactions at the molecular scale, whose sheer size
  obfuscates understanding. In light of limited experimental data,
  many parameters of such dynamics are unknown, and thus models built
  on the detailed, mechanistic viewpoint overfit and are not
  predictive. At the other extreme, simple {\em ad hoc} models of
  complex processes often miss defining features of the underlying
  systems. Here we propose an approach that instead constructs {\em
    phenomenological}, coarse-grained models of network dynamics that
  automatically adapt their complexity to the amount of available
  data. Such adaptive models lead to accurate predictions even when
  microscopic details of the studied systems are unknown due to
  insufficient data.  The approach is computationally tractable, even
  for a relatively large number of dynamical variables, allowing its
  software realization, named {\em Sir Isaac}, to make successful
  predictions even when important dynamic variables are
  unobserved. For example, it matches the known phase space structure
  for simulated planetary motion data, avoids overfitting in a complex
  biological signaling system, and produces accurate predictions for a
  yeast glycolysis model with only tens of data points and over half
  of the interacting species unobserved.\\
  
  $^\ast$E-mail: bdaniels@discovery.wisc.edu,
    ilya.nemenman@emory.edu

\end{abstract}

\maketitle



\begin{bibunit}[unsrt]

\section{Introduction}

Systems biology is a field of complicated models --- and rightfully
so: the vast amount of experimental data has clearly demonstrated that
cellular networks have a degree of complexity that is far greater than
what is normally encountered in the physical world
\cite{Hlavacek:2009kf}. Mathematical models of these data are often as
complicated as the data themselves, reflecting the humorous maxim that
``the best material model of a cat is another, or preferably the same,
cat'' \cite{Rosenblueth:1945tz}. However, continued success of
approaches that systematize all known details in a combinatorially
large mathematical model is uncertain. Indeed, generalizing and
generating insight from complex models is difficult. Further,
specification of myriads of microscopic mechanistic parameters in such
models demands vast data sets and computational resources, and
sometimes is impossible even from very large data sets due to widely
varying sensitivities of predictions to the parameters
\cite{Gutenkunst:2007gl}.  Finally, the very structures of these
models are often unknown because they depend on many yet-unobserved
players on the cellular, molecular, and sub-molecular levels.
Identification of these structural characteristics of the involved
processes is labor intensive and does not scale up easily.  With these
challenges, it is unlikely that mathematical models based solely on a
reductionist representation will be able to account accurately for the
observed dynamics of cellular networks. More importantly, even if they
could, the resulting models would be too unwieldy to bring about
{\em understanding} of the modeled systems.

Because of these difficulties, the need to use systems biology data to
predict responses of biological systems to dynamical perturbations,
such as drugs or disease agents, has led to a resurgence of research
into automated inference of dynamical systems from time series data,
which had been attempted since the early days of the field of
nonlinear dynamics \cite{Crutchfield:1987va,PacCruFar80}.  Similar
needs in other data-rich fields in natural and social sciences and
engineering have resulted in successful algorithms for distilling
continuous dynamics from time series data, using approaches such as
linear dynamic models \cite{FriHarPen03}, recurrent neural networks
\cite{SusAbb09}, evolved regulatory networks \cite{FraHakSig07}, and
symbolic regression \cite{Schmidt:2009dt,SchValJen11}. The latter two
approaches produce models that are more mechanistically interpretable
in that they incorporate nonlinear interactions that are common in
systems biology, and they actively prune unnecessary complexity.  Yet
these approaches are limited because, in a search through all possible
microscopic dynamics, computational effort typically explodes with the
growing number of dynamical variables.  In general, this leads to very
long search times \cite{FraHakSig07,SchValJen11}, especially if some
underlying variables are unobserved, and dynamics are coupled and
cannot be inferred one variable at a time.

To move forward, we note that, while biological networks are complex,
they often realize rather simple input-output relations, at least in
typical experimental setups. Indeed, activation dynamics of a
combinatorially complex receptor can be specified with only a handful
of large-scale parameters, including the dynamic range, cooperativity,
and time delay \cite{Goldstein:2004ch,Bel:2010er,Cheong:2011jp}. Also,
some microscopic structural complexity arises in order to guarantee
that the macroscopic functional output remains simple and robust in
the face of various perturbations \cite{Bel:2010er,Lander:2011bj}.
Thus one can hope that macroscopic prediction does not require
microscopic accuracy \cite{MacChaTra13}, and hence seek
phenomenological, coarse-grained models of cellular processes that are
simple, inferable, and interpretable, and nonetheless useful in
limited domains.


In this report, we propose an adaptive approach for dynamical
inference that does not attempt to find the single best
microscopically ``correct'' model, but rather a phenomenological model
that remains mechanistically interpretable and is ``as simple as
possible, but not simpler'' than needed to account for the
experimental data. Relaxing the requirement for microscopic accuracy
means that we do not have to search through all possible microscopic
dynamics, and we instead restrict our search to a much smaller
hierarchy of models.  By choosing a hierarchy that is nested and
complete, we gain theoretical guarantees of statistical consistency,
meaning the approach is able to adaptively fit any smooth dynamics
with enough data, yet is able to avoid problems with overfitting that
can happen without restrictions on the search space
\cite{Nem05}. While similar complexity control methods are well
established in statistical inference \cite{MacKay:2003wc}, we believe
that they have not been used yet in the context of inferring complex,
nonlinear {\em dynamics}.  Importantly, this adaptive approach is
typically much more efficient because there are far fewer models to
test.  Instead of searching a super-exponentially large model space
\cite{Schmidt:2009dt}, our method tests a number of models that scales
polynomially with the number of dynamical variables.  Further, it uses
computational resources that asymptotically scale linearly with the
number of observations.  This allows us to construct interpretable
models with much smaller computational effort and fewer experimental
measurements, even when many dynamical variables are unobserved. We
call the approach {\em Sir Isaac} due to its success in discovering the
law of universal gravity from simulated data (see below).


\section{Methods and Results}

\subsection{Classes of phenomenological models used by Sir Isaac}

We are seeking a phenomenological model of dynamics in the form
\begin{equation}
\frac{d\vec{x}}{dt}=\vec{F_x}(\vec{x},\vec{y},\vec{I}),
~~
\frac{d\vec{y}}{dt}=\vec{F_y}(\vec{x},\vec{y},\vec{I}),
\label{eq:model}
\end{equation}
where $\vec{x}$ are the observed variables, $\vec{y}$ are the hidden
variables, and $\vec{I}$ are the inputs or other parameters to the
dynamics. We neglect intrinsic stochasticity in the dynamics (either
deterministic chaotic, or random thermal), and focus on systems where
repeated observations with nearly the same initial conditions produce
nearly the same time series, save for measurement noise. The goal is
then to find a phenomenological model of the force fields
$\vec{F}_x,\vec{F}_y$ \cite{Crutchfield:1987va}. The same dynamics may
produce different classes of trajectories $\vec{x}(t)$ dependent on
initial conditions (e.~g., elliptical vs.\ hyperbolic trajectories in
gravitational motion). Thus the focus on {\em dynamical} inference
rather than on more familiar statistical modeling of trajectories
allows the representation of multiple functional forms within a single
dynamical system.

To create a model, we would like to gradually increase the complexity
of $F$ until we find the best tradeoff between good fit and sufficient
robustness, essentially extending traditional Bayesian model selection
techniques to the realm of dynamical models.  Ideally, this process
should progress much like a Taylor series approximation to a function,
adding terms one at a time in a hierarchy from simple to more complex,
until a desired performance is obtained. To guarantee that this is
possible, the hierarchy of models must be nested (or ordered) and
complete in the sense that any possible dynamics can be represented
within the hierarchy \cite{Nem05} (see {\em Supplementary Online
  Materials (SOM)}).  Any model hierarchy that fits these criteria may
be used, but ordering dynamical models that can be made more complex
along two dimensions (by adding either nonlinearities or unobserved
variables) is nontrivial. Further, different model hierarchies may
naturally perform differently on the same data, depending on whether
the studied dynamics can be represented succinctly within a hierarchy.

We construct two classes of nested and complete model hierarchies,
both well matched to properties of biochemistry that underlies
cellular network dynamics. We build the first with S-systems
\cite{SavVoi87} and the second with continuous time sigmoidal networks
\cite{Bee06} (see {\em SOM}).  The S-systems use production and
degradation terms for each dynamical variable formed by products of
powers of species concentrations; this is a natural generalization of
biochemical mass-action laws. The sigmoidal class represents
interactions using linear combinations of saturating functions of
species concentrations, similar to saturation in biochemical reaction
rates.  Both classes are complete and are able to represent any smooth
dynamics with a sufficient number of (hidden) dynamical variables
\cite{SavVoi87,FunNak93,ChoLi00}. It is possible that both classes can
be unified into power-law dynamical systems with algebraic power-law
constraints among the dynamical variables \cite{SavVoi87}, but this
will not be explored in this report.

\subsection{Description of model selection procedure}
To perform adaptive fitting within a model class, a specific ordered
hierarchy of models is chosen \textit{a priori} that simultaneously
varies both the degree of nonlinearity and the number of hidden
variables (see \figref{modelSpaceDiagram} and {\em SOM}).  For each
model in the hierarchy, its parameters are fit to the data and an
estimate of the Bayesian log-likelihood $\L$ of the model is
calculated.  This estimate makes use of a generalized version of the Bayesian
Information Criterion \cite{Sch78}, which we have adopted, for the
first time, for use with nonlinear dynamical systems inference.  As
models increase in complexity, $\L$ first grows as the quality of fit
increases, but eventually begins to decrease, signifying
overfitting. Since, statistical fluctuations aside, there is just one
peak in $\L$ \cite{Nem05}, one can be certain that the global maximum
has been observed once it has decreased sufficiently. The search
through the hierarchy is then stopped, and the model with maximum $\L$
is ``selected'' (see \figref{yeastModelSelection}(b)).

\subsection{The law of gravity}

\begin{figure}
\centering
\includegraphics[scale=1.25]{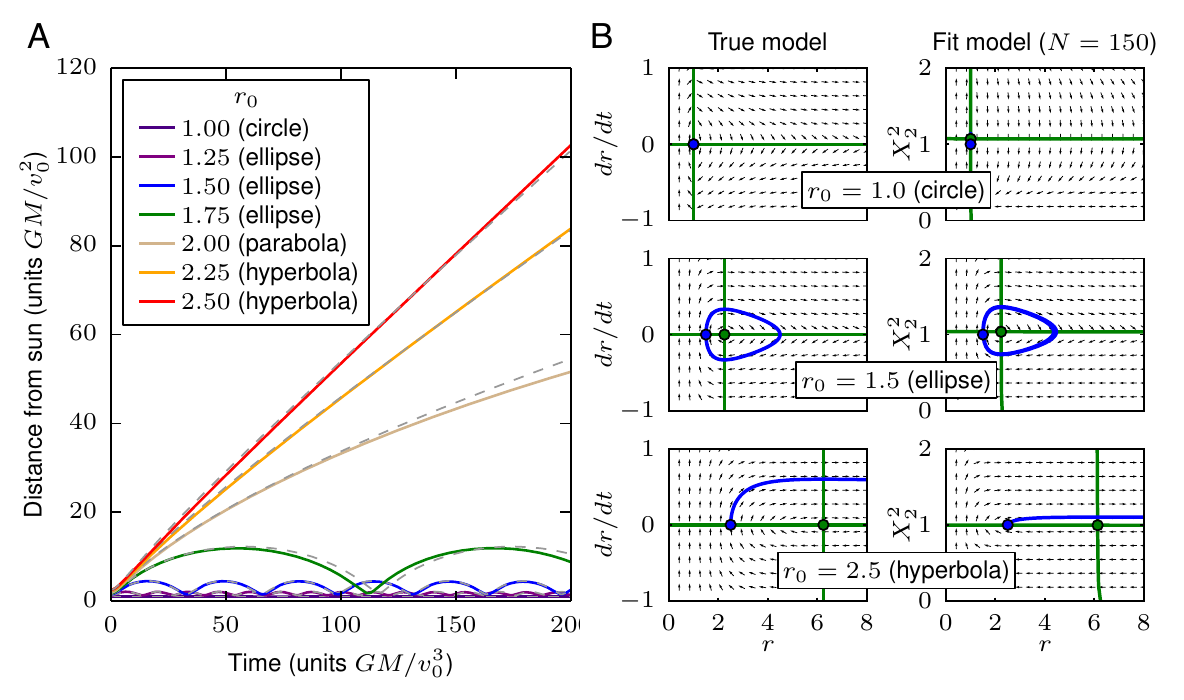}
\caption{\label{planetary} The law of gravity: an example of dynamical
  inference.  A particle is released with velocity $v_0$ perpendicular
  to the line connecting it to the sun, with varying initial distance
  $r_0$ from the sun.  (a) With only $N = 150$ examples (each
  consisting of just a single noisy observation of $r$ at a random
  time $t$ after the release; see {\em SOM}), we infer a single
  dynamical model in the S-systems class that reproduces the data.
  With no supervision, adaptive dynamical inference produces
  bifurcations that lead to qualitatively different behavior: in this
  case, a single model produces both oscillations (corresponding to
  elliptical orbits) and monotonic growth (corresponding to hyperbolic
  trajectories).  Inferred trajectories are shown with solid colored
  lines, and the corresponding true trajectories are shown with dashed
  lines.  (b) Like the true model (left), the inferred model (right)
  contains a single hidden variable $X_2$ and works using a similar
  phase space structure.  Specifically, the location of nullclines
  (green lines) and a single fixed point (green circle) as a function
  of $r_0$ are recovered well by the fit. Note that the hidden
  variable is defined up to a power (see {\em SOM}), and we choose to
  plot $X^2_2$ here. }
\end{figure}

Before applying the approach to complex biological dynamics, where the
true model may not be expressible simply within the chosen search
hierarchy, we test it on a simpler system with a known exact
solution. We choose the iconic law of gravity, inferred by Newton
based on empirical observations of trajectories of planets, the Moon,
and, apocryphally, a falling apple. Crucially, the
inverse-squared-distance law of Newtonian gravity can be represented
exactly within the S-systems power-law hierarchy for elliptical and
hyperbolic trajectories, which do not go to zero radius in finite
time. It requires a hidden parameter, the velocity, to completely
specify the dynamics of the distance of an object from the sun (see
{\em SOM} for specification of the model).

\figref{planetary} displays the result of the adaptive inference using
the S-systems class. When given data about distance of an object from
the sun over time, we discover a model that reproduces the underlying
dynamics, including the necessary hidden variable and the bifurcation
points. Since the trajectories include hyperbolas and ellipses, this
example emphasizes the importance of inferring a single set of
dynamical equations of motion, rather than statistical fits to
trajectories themselves, which would be different for the two cases.
\figref{planetarySigmoidal} additionally shows fits for the law of
gravity using the sigmoidal models class. While accurate, the fits are
worse than those for the S-systems, illustrating importance of
understanding of basic properties of the studied system when
approaching automated model inference. 

Empowered by the success of the adaptive inference approach for this
problem, we chose to name it {\em Sir Isaac}. The software
implementation can be found under the same name on GitHub.

\subsection{Multi-site phosphorylation model}

When inferring models for more general systems, we do not expect the
true dynamics to be perfectly representable by any specific model
class: even the simplest biological phenomena may involve
combinatorially many interacting components.  Yet for simple
macroscopic behavior, we expect to be able to use a simple approximate
model that can produce useful predictions.  To demonstrate this, we
envision a single immune receptor with $n$
modification sites, which can exist in $2^n$ microscopic states
\cite{HlaFaeBli06}, yet has simple macroscopic behavior for many
underlying parameter combinations.  Here, we test a model receptor
that can be phosphorylated at each of $n=5$ sites arranged in a linear
chain.  The rates of phosphorylation and dephosphorylation at each
site are affected by the phosphorylation states of its nearest
neighboring sites.  This produces a complicated model with 32 coupled
ODEs specified by 52 parameters, which we assume are unknown to the
experimenter.

We imagine an experimental setup in which we can control one of these
parameters, and we are interested in its effects on the time evolution
of the total phosphorylation of all 5 sites.  Here, we treat as input
$I$ the maximum rate of cooperative phosphorylation of site 2 due to
site 3 being occupied, $V$, and measure the resulting time course of
total phosphorylation starting from the unphosphorylated state.
Experimental measurements are corrupted with noise at the
scale of 10\% of their values ({\em SOM}).

\begin{figure}
\centering
\includegraphics[scale=1.0]{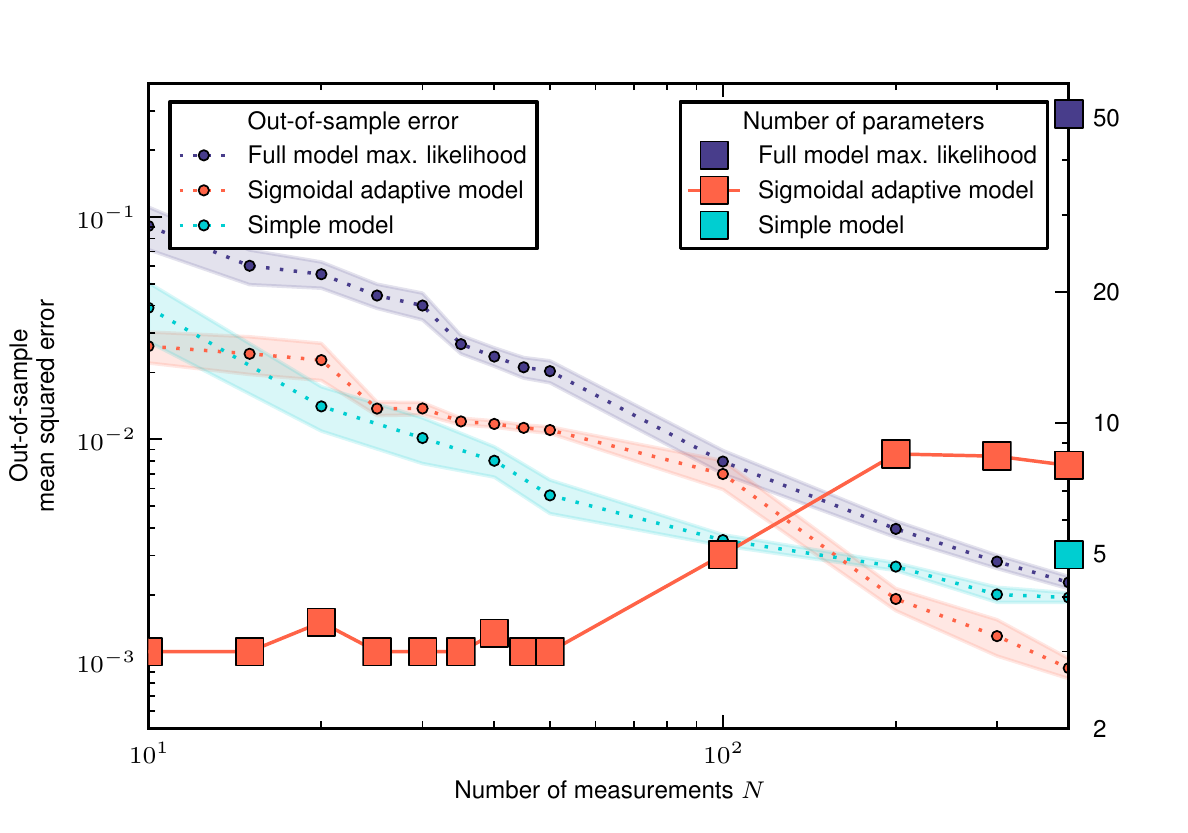}
\caption{\label{phosNdependence}
  Multi-site phosphorylation model selection as a function of the
  number of measurements $N$.  The sizes of errors made by three models
  decrease as the amount of data increases.  Adaptive sigmoidal models
  perform roughly as well as a custom-made simple 5-parameter model
  for small $N$, but outperform the simple model for large amounts of
  data.  Although we expect that it will eventually outperform all
  other models as $N \rightarrow \infty$, a maximum likelihood fit to
  the full 52-parameter model (dark blue) performs worse in this range
  of $N$.  The mean over 10 sets of input data are shown, with shaded
  regions indicating the standard deviation of the mean.  On the right axis,
  the number of parameters in each model is indicated, with the
  sigmoidal model adapting to use more parameters when given more data
  (red squares).  }
\end{figure}

A straightforward approach to modeling this system is to fit the 52
parameters of the known model to the phosphorylation data.  A
second approach is to rely on intuition to manually develop a
functional parameterization that captures the most salient features of
the timecourse data.  In this case, a simple 5 parameter model (see
{\em SOM}) captures exponential saturation in time with an asymptotic
value that depends sigmoidally on the input $V$.  A third approach,
advocated here, is to use automated model selection to create a model
with complexity that matches the amount and precision of the available
data.

\begin{figure}
\centering
\includegraphics[scale=1.25]{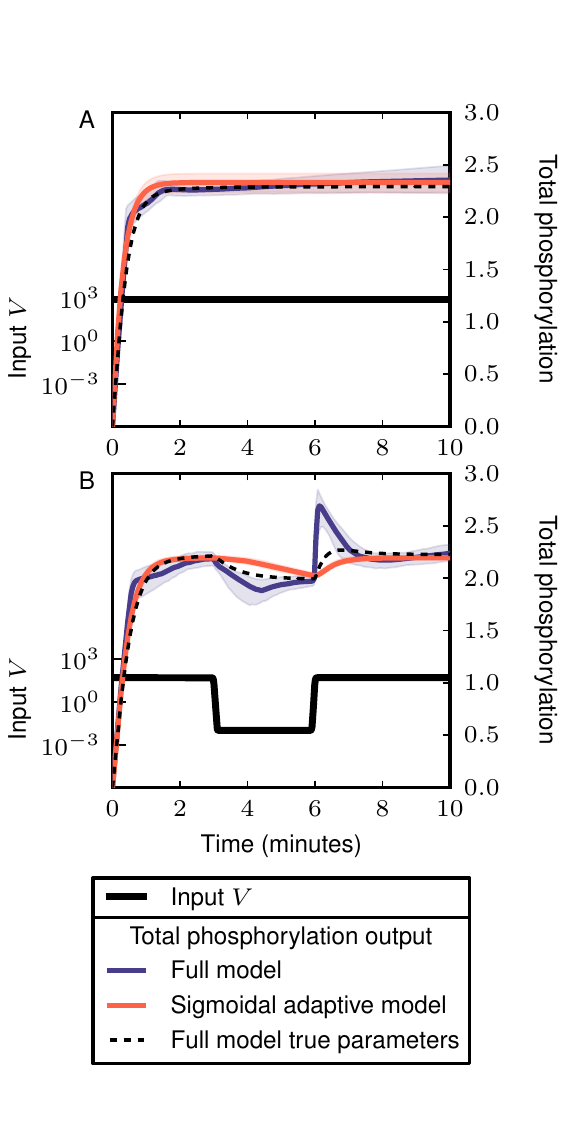}
\caption{\label{phosTimecourse}
  Response (right axis) to (a) out-of-sample constant and (b)
  time-varying input (left axis) in the models of multi-site
  phosphorylation.  Fit to $N = 300$ constant input data points, 
  the full known model (dark blue) produces erratic
  behavior typical of overfitting, while the adaptive sigmoidal model
  (red) produces more stable out-of-sample predictions with median
  behavior that is closer to the true dynamics.  Dark lines indicate
  the median behavior over 100 samples from each model's parameter
  posterior (see {\em SOM}), and shaded regions indicate 90\%
  confidence intervals.  }
\end{figure}

In \figref{phosNdependence}, we compare these three approaches as the
amount of available data is varied, and
\figref{phosTimecourse}(a)
shows samples of fits done by different procedures.  With
limited and noisy data, fitting the parameters of the full known model
risks overfitting, and in the regime we test, it is the worst
performer on out-of-sample predictions.  The simple model performs
best when fitting to less than 100 data points, but for larger amounts
of data it saturates in performance, as it cannot fit more subtle
effects in the data.  In contrast, an adaptive model remains simple
with limited data and then grows to accommodate more subtle behaviors
once enough data is available, eventually outperforming the simple
model.

The multi-site phosphorylation example also demonstrates that
dynamical phenomenological models found by {\em Sir Isaac} are more
than fits to the existing data, but rather they uncover the true
nature of the system in a precise sense: they can be used to make
predictions of model responses to some classes of inputs that are
qualitatively different from those used in the inference. For example,
as seen in \figref{phosTimecourse}(b), an adaptive sigmoidal model
inferred using {\em temporally constant} signals produces a reasonable
extrapolated prediction for response to a {\em time-varying}
signal. At the same time, overfitting is evident when using the full,
detailed model, even when one averages the model responses over the
posterior distribution of the inferred model parameters.

\subsection{Yeast glycolysis model}

\begin{figure}
\centering
\includegraphics[scale=1.2]{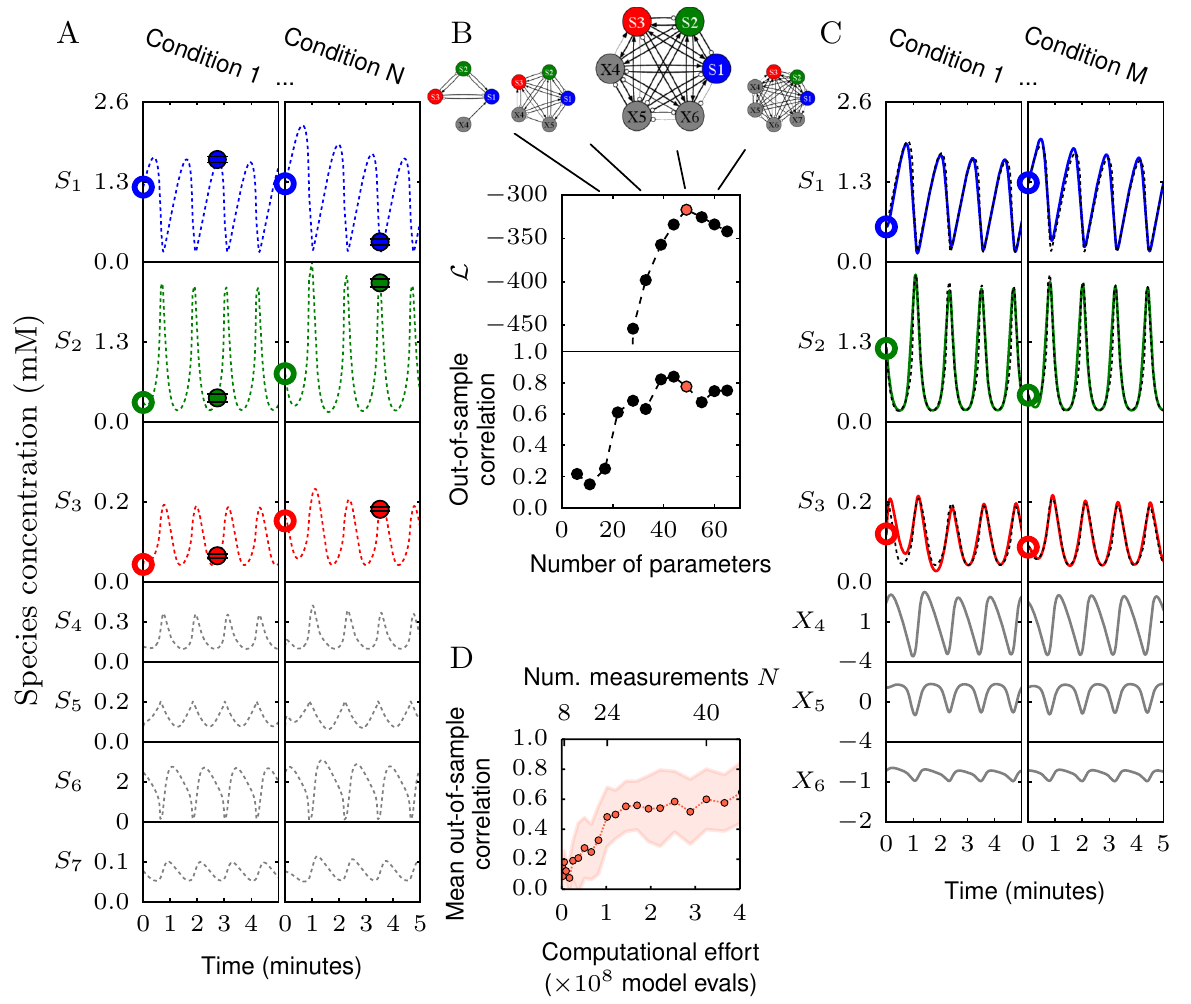}
\caption{\label{yeastModelSelection}
  An example of the model selection process using measurements of
  timecourses of three metabolites in yeast glycolysis as their
  initial concentrations are varied.  (a) For each set of initial
  conditions (open circles), a noisy measurement of the three
  observable concentrations (filled circles) is made at a single
  random time.  Hidden variables (in gray) are not measured.  In this
  example, we fit to $N = 40$ in-sample conditions.  (b)
  Models from an ordered class, with the illustrated connectivity, are
  fit and tested sequentially until $\L$, an approximation of the
  relative log-likelihood, decreases sufficiently from a maximum.
  (c) The selected model (large connectivity diagram) is used to
  make predictions about out-of-sample conditions.  Here, we compare
  the output of the selected model (solid lines) to that of the model
  that created the synthetic data (dashed lines).  (d)
  Performance versus computational and experimental effort.  The mean
  out-of-sample correlation for 3 measured biochemical species from
  the range of initial conditions twice that used in training rises to
  over 0.6 using less than $5 \times 10^8$ model evaluations and 40
  in-sample measurements.  In Ref.~\cite{SchValJen11}, inferring an
  exact match to the original 7-dimensional model used roughly 500
  times as many measurements of all 7 species (with none hidden),
  which were chosen carefully to be informative. The approach also
  uses 200 times as many model evaluations (see {\em
    SOM}). Nonetheless, the accuracy of both approaches is comparable,
  and {\em Sir Isaac} additionally retains information about the phase
  of the oscillations.}
\end{figure}

A more complicated system, for which there has been recent interest in
automated inference, is the oscillatory dynamics of yeast glycolysis
\cite{SchValJen11}.  A recent model for the system
\cite{WolHei00,RuoChrWol03}, informed by detailed knowledge of
cellular metabolic pathways, consists of coupled ODEs for 7 species
with concentrations that oscillate with a period of about 1
minute. The system dynamic is simpler than its structure in the sense
that some of the complexity is used to stabilize the oscillations to
external perturbations. On the other hand, the oscillations are not
smooth (see \figref{yeastModelSelection}) and hence are hard to fit
with simple methods. These considerations make this model an ideal
next test case for phenomenological inference with {\em Sir
  Isaac}.

If we were given abundant time series data from all 7 species and were
confident that there were no other important hidden species, we may be
in a position to infer a ``true'' model detailing interactions among
them.  If we are instead in the common situation of having limited
data on a limited number of species, we may more modestly attempt to
make predictions about the types of inputs and outputs that we have
measured.  This is conceptually harder since an unknown number of
hidden variables may need to be introduced to account for the dynamics
of the observed species. We demonstrate our approach by constructing
adaptive models of the dynamics from data for only 3 of the 7 coupled
chemical species, as their initial conditions are varied.

Depicted in \figref{yeastModelSelection} is the model selection
procedure for this case.  After selecting an adaptive model fit to
noisy data from $N$ single timepoints, each starting from initial
conditions sampled from specified ranges, we test the inferred model's
ability to predict the timecourse resulting from out-of-sample initial
conditions.  With data from only $N = 40$ measurements, the selected
model is able to predict behavior with mean correlation of over 0.6
for initial conditions chosen from ranges {\em twice as large} as
those used as training data (shown in \figref{yeastModelSelection})
and 0.9 for out-of-sample ranges equal to in-sample ranges (shown in
\figref{yeastVsN}).  Previous work that inferred the exact equations
of the original 7-dimensional model \cite{SchValJen11} used roughly
500 times as many measurements of all 7 variables and 200 times as
many model evaluations.  This example also demonstrates that adaptive
modeling can hint at the complexity of the hidden dynamics beyond
those measured: the best performing sigmoidal model requires three
hidden variables, for a total of six chemical species --- only one less
than the true model.  Crucially, the computational complexity of {\em
  Sir Isaac} still scales linearly with the number of observations,
even when a large fraction of variables remains hidden (see {\em SOM}
and \figref{modelEvaluations}).

\section{Discussion}
The three examples demonstrate the power of the adaptive,
phenomenological modeling approach.  {\em Sir Isaac} models are
inferred without an exponentially complex search over model
space, which would be impossible for systems with many
variables. These dynamical models are as simple or complex as
warranted by data and are guaranteed not to overfit even for small
data sets. Thus they require orders of magnitude less data and
computational resources to achieve the same predictive accuracy as
more traditional methods that infer a pre-defined, large number of
mechanistic parameters in the true model describing the system.

These advantages require that the inferred models are
phenomenological, and are designed for efficiently predicting the
system dynamics at a given scale, determined by the available
data. While \figref{planetary} shows that {\em Sir Isaac} will infer
the true model if it falls within the searched model hierarchy, and
enough data is available, more generally, the inferred dynamics may be
quite distinct from the true microscopic, mechanistic processes, as
shown by a different number of chemical species in the true and the
inferred dynamics in \figref{yeastModelSelection}. What is then the
utility of the approach if it says little about the underlying
mechanisms?

First, there is the obvious advantage of being able to predict
responses of systems to yet-unseen experimental conditions, including
those qualitatively different from the ones used for
inference. Second, some general mechanisms, such as the necessity of
feedback loops or hidden variables, are easily uncovered even in
phenomenological models. However, more importantly, we draw the
following analogy. When in the 17th century Robert Hooke studied the
force-extension relations for springs, a linear model of the relation
for a specific spring did not tell much about the mechanisms of force
generation. However, the observation that {\em all} springs exhibit
such linear relations for small extensions allowed him to combine the
models into a law --- Hooke's law, the first of many phenomenological
physical laws that followed. It instantly became clear that
experimentally measuring just one parameter, the Hookean stiffness,
provided an exceptionally precise description of the spring's
behavior. And yet the mechanistic understanding of how this Hooke's
constant is related to atomic interactions within materials is only
now starting to emerge. Similarly, by studying related phenomena
across complex biological systems (e.g., chemotactic behavior in {\em
  E.~coli} \cite{Berg:2004wy} and {\em C.~elegans} \cite{Ryu02}, or
behavioral bet hedging, which can be done by a single cell
\cite{Kussell:2005dg} or a behaving rodent \cite{Gallistel:2001wi}),
we hope to build enough {\em models} of specific systems, so that
general {\em laws} describing how nature implements them become
apparent.

If successful, our search for phenomenological, emergent dynamics
should allay some of the most important skepticism regarding the
utility of automated dynamical systems inference in science
\cite{Anderson:2009kc}, namely that such methods typically start with
known variables of interest and known underlying physical laws, and
hence cannot do transformative science and find new laws of
nature. Indeed, we demonstrated that, for truly successful
predictions, the model class used for automated phenomenological
inference must match basic properties of the studied dynamics
(contrast, for example, \figref{planetary} to
\figref{planetarySigmoidal}, and see
\figref{phosTimecourse_appendix}).  Thus fundamental understanding of
some key properties of the underlying mechanisms, such as the
power-law structure of the law of gravity, or the saturation of
biochemical kinetic rates, can be inferred from data even if unknown
{\em a priori}.  Finally, we can contrast our approach with a standard
procedure for producing coarse-grained descriptions of inanimate
systems: starting from a mechanistically accurate description of the
dynamics, and then mapping them onto one of a small set of
universality classes \cite{Wilson,MacChaTra13}.  This procedure is
possible due to symmetries of physical interactions that are not
typically present in living systems.  Without such symmetries, the
power of universality is diminished, and microscopic models may result
in similarly different macroscopic ones. Then specifying the
microscopic model first in order to coarse-grain it later becomes an
example of solving a harder problem to solve a simpler one
\cite{Vapnik}.  Thus for living systems, the direct inference of
phenomenological dynamics, such as done by {\em Sir Isaac}, may be the
optimal way to proceed.



\putbib[PredictivePhenomenology]

\end{bibunit}

\begin{acknowledgments}
We thank William Bialek and Michael Savageau for important
  discussions, Andrew Mugler, David Schwab, and Fereydoon Family for
  their critical comments, and the hospitality of the Center for
  Nonlinear Studies at Los Alamos National Laboratory. This research
  was supported in part by the James S.\ McDonnell foundation Grant
  No. 220020321 (I.\ N.), a grant from the John Templeton Foundation
  for the study of complexity (B.\ D.), the Los Alamos National
  Laboratory Directed Research and Development Program (I.\ N.\ and
  B.\ D.), and NSF Grant No. 0904863 (B.\ D.).
\end{acknowledgments}


\begin{bibunit}[unsrt]

\clearpage

\renewcommand{\thepage}{S\arabic{page}}
\setcounter{page}{1}
\renewcommand{\theequation}{S\arabic{equation}}
\setcounter{equation}{0}
\renewcommand{\thefigure}{S\arabic{figure}}
\setcounter{figure}{0}
\renewcommand{\thetable}{S\arabic{table}}
\setcounter{table}{0}
\renewcommand{\thesection}{S}
\setcounter{section}{0}
\setcounter{subsection}{0}


\section*{Supplementary Materials}

\subsection{Materials and Methods}

\subsubsection{Hierarchical Bayesian model selection}

\begin{figure}
\centering
\includegraphics[scale=0.4]{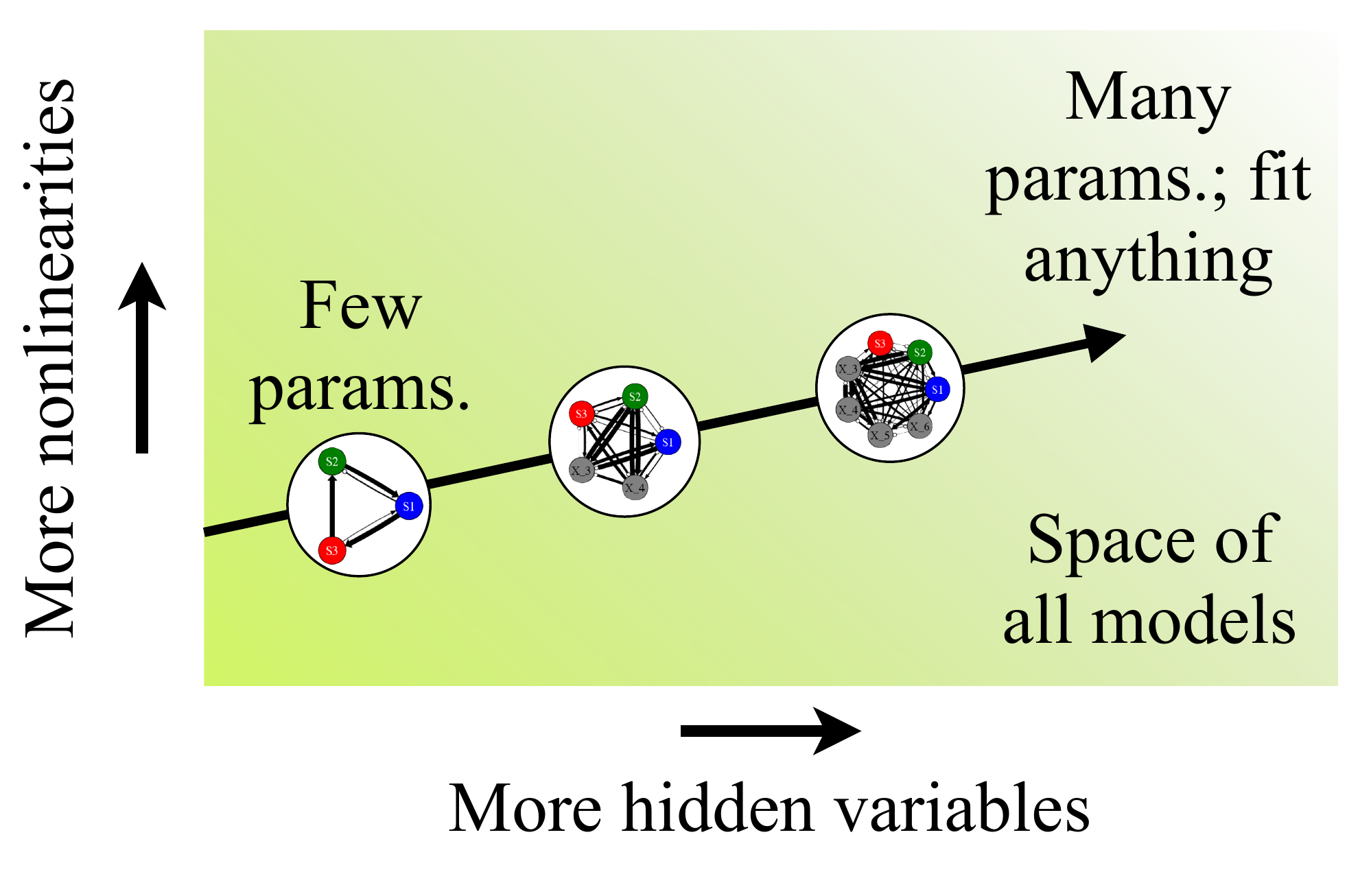}
\caption{\label{modelSpaceDiagram}
Hierarchical model selection follows a single predefined
path through model space. 
}
\end{figure}

For consistent inference, we need a hierarchy of models that satisfies
criteria laid out in Ref.~\cite{Nem05}.  First, we desire a model
hierarchy that will produce a single maximum in $\L$, up to
statistical fluctuations, as we add complexity. For this, the
hierarchy should be nested (but not necessarily regular or
self-similar), meaning that once a part of the model is added, it is
never taken away.  Second, the hierarchy should be complete, meaning
it is able to fit any data arbitrarily well with a sufficiently
complex model.  Intuitively, instead of searching a large
multidimensional space of models, hierarchical model selection follows
a single predefined path through model space
(\figref{modelSpaceDiagram}).  While the predefined path may be
suboptimal for a particular instance (that is, the true model may not
fall on it), even then the completeness guarantees that we will still
eventually learn any dynamical system $F$ given enough data, and
nestedness assures that this will be done without overfitting along
the way.\footnote{ In general, we are not guaranteed good predictive
  power until $N \rightarrow \infty$, but we can hope that the
  assumptions implicit in our priors (consisting of the specific form
  of the chosen model hierarchy and the priors on its parameters) will
  lead to good predictive power even for small $N$.}

\subsubsection{Adaptive model classes and hierarchies}

Our first model class is the S-system power-law class.  The general
form of the S-system representation consists of $J$ dynamical
variables $x_i$ and $K$ inputs $I_k=x_{J+k}$, with each dynamical
variable governed by an ordinary differential equation:
\cite{SavVoi87} 
\begin{equation}
\label{generalPowerlaw}
\frac{d x_i}{dt} = G(\vecx)_i - H(\vecx)_i,
\end{equation}
with production $G$ and degradation $H$ of the form
\begin{eqnarray}
G(\vecx)_i &=& \alpha_i \prod_{j=1}^{J+K} x_j^{g_{ij}} \\
H(\vecx)_i &=& \beta_i  \prod_{j=1}^{J+K} x_j^{h_{ij}}.
\end{eqnarray}
In a process called ``recasting,'' any set of differential equations
written in terms of elementary functions can be rewritten in the
power-law form by defining new dynamical variables in the correct way
\cite{SavVoi87}.  Since any sufficiently smooth function can be
represented in terms of a series of elementary functions (e.\ g.,
Taylor series), a power-law network of sufficient size can
describe any such deterministic dynamical system.  Note that, 
since exponents are
not constrained to be positive or integer-valued, dynamics in this
class are generally ill-defined when variables are not positive.

We find that the S-systems model class works well for planetary
motion, which has an exact representation in the class; see
Section~\ref{planetarySection}.  For our biological test examples, the
S-systems class is outperformed by the sigmoidal class (see below).
This may be indicating that behavior common in the S-systems class is
not common in typical biological systems (e.\ g., real production and
degradation terms cannot grow without bounds). It may also stem from
the positivity constraint: since the condition that variables remain
positive is not easily determined from parameter values, we are forced
in our model selection process to simply discard any tested parameters
that lead to zero or negative values.

The second model hierarchy is the sigmoidal network class.  In this
class, we use the fact that the interactions among biological
components often take the form of a sigmoidal function to define the
following system of ODEs:
\begin{equation}
\frac{d x_i}{dt} = 
    -x_i/\tau_i + \sum_{j=1}^J W_{ij} ~ \xi(x_j + \theta_j)
    + \sum_{k=1}^{N_p} V_{ik} \Input_k,
\end{equation}
where the sigmoidal function $\xi(y) = 1/(1+e^{y})$.  This class of
models has also been shown to approximate any smooth dynamics
arbitrarily well with a sufficient number of dynamical variables
\cite{Bee06,BeeDan10,FunNak93,ChoLi00}. Note that natural variations
of this class to be explored in future work include rescaling of the
arguments of the sigmoids $\xi$ or switching the order of operations
to apply the sigmoidal function to a linear combination of state
variables in order to more closely match traditional neural network
models \cite{RumHinWil86}.

An advantage of the S-systems and sigmoidal representations is the
existence of a natural scheme for creating a one-dimensional model
hierarchy: simply adding dynamical variables $x_i$.  The most general
network is fully connected, such that every variable $x_i$ has an
interaction term in every other $d x_j/dt$.  Our hierarchy starts with
a fully-connected network consisting of the necessary number of input
and output variables, and adds ``hidden'' dynamical variables to add
complexity.  With each additional $x_i$, we add parameters in a
predetermined order.

In the S-systems class, without connections, variable $x_i$'s behavior
is specified by 5 parameters: $x_{i}^{\rm
  init},\alpha_i,\beta_i,g_{ii}$, and $h_{ii}$.  Each connection to and
from $x_j$ is specified by 4 parameters: $g_{ij},g_{ji},h_{ij},$ and
$h_{ji}$.  When adding a new dynamic variable, we first fix its
parameters (to zero for the exponential parameters and one for the
multiplicative parameters), and then allow them to vary one at a time
in the following order:
$g_{ii},g_{ji},h_{ji},g_{ij},h_{ij},\beta_i,h_{ii},\alpha_i$ (adding
connections to every other $x_j$ one at a time).  An example is shown
in Table~\ref{powerLawTable}.

\renewcommand*\arraystretch{1.5}
\begin{table}
\centering
\begin{tabular}{|c c l|}
  \hline Model No.\ $i$ & Num. parameters $N_p$    
  &  ~Form of power-law ODEs  \\ \hline
  0 & 3 & $\begin{aligned}[t]
    x_1(0) &= x_1^{\rm init} \\
    \frac{dx_1}{dt}&= x_I^{g_{1 0}} x_1^{g_{1 1}}
    - \beta_1 
    \end{aligned}$ \\[30pt] \hline
    1 & 4 & $\begin{aligned}[t]
        x_1(0) &= x_1^{\rm init} \\
        \frac{dx_1}{dt}&= x_I^{g_{1 0}} x_1^{g_{1 1}}
            - \beta_1 x_I^{h_{1 0}} 
    \end{aligned}$ \\[30pt] \hline
    2 & 5 & $\begin{aligned}[t]
        x_1(0) &= x_1^{\rm init} \\
        \frac{dx_1}{dt}&= x_I^{g_{1 0}} x_1^{g_{1 1}}
            - \beta_1 x_I^{h_{1 0}} x_1^{h_{1 1}}
    \end{aligned}$ \\[30pt] \hline
    3 & 6 & $\begin{aligned}[t]
        x_1(0) &= x_1^{\rm init} \\
        \frac{dx_1}{dt}&= \alpha_1 x_I^{g_{1 0}} x_1^{g_{1 1}}
            - \beta_1 x_I^{h_{1 0}} x_1^{h_{1 1}} \\
    \end{aligned}$ \\[30pt] \hline
    4 & 8 & $\begin{aligned}[t] 
        x_1(0) &= x_1^{\rm init} \\
        x_2(0) &= x_2^{\rm init} \\
        \frac{dx_1}{dt}&= \alpha_1 x_I^{g_{1 0}} x_1^{g_{1 1}} x_2^{g_{1 2}}
            - \beta_1 x_I^{h_{1 0}} x_1^{h_{1 1}} \\
        \frac{dx_2}{dt}&= x_2^{g_{2 2}} - 1
    \end{aligned}$ \\[76pt] \hline
    5 & 9 & $\begin{aligned}[t] 
        x_1(0) &= x_1^{\rm init} \\
        x_2(0) &= x_2^{\rm init} \\
        \frac{dx_1}{dt}&= \alpha_1 x_I^{g_{1 0}} x_1^{g_{1 1}} x_2^{g_{1 2}}
            - \beta_1 x_I^{h_{1 0}} x_1^{h_{1 1}} x_2^{h_{1 2}} \\
        \frac{dx_2}{dt}&= x_2^{g_{2 2}} - 1
    \end{aligned}$ \\[76pt] \hline
    6 & 10 & $\begin{aligned}[t] 
        x_1(0) &= x_1^{\rm init} \\
        x_2(0) &= x_2^{\rm init} \\
        \frac{dx_1}{dt}&= \alpha_1 x_I^{g_{1 0}} x_1^{g_{1 1}} x_2^{g_{1 2}}
            - \beta_1 x_I^{h_{1 0}} x_1^{h_{1 1}} x_2^{h_{1 2}} \\
        \frac{dx_2}{dt}&= x_1^{g_{2 1}} x_2^{g_{2 2}} - 1
    \end{aligned}$ \\[76pt] \hline
\end{tabular}
\caption{\label{powerLawTable}
  The first seven models of an 
  example hierarchy in the S-systems class
  with one input $x_I$ and fixed initial conditions $x_1^{\rm init}$ and
  $x_2^{\rm init}$.
}
\end{table}


The sigmoidal class is similar:
without connections, variable $x_i$'s behavior is specified by 4 parameters:
$x_{i}^{init},W_{ii},\tau_i$, and $\theta_i$.  
Each connection to and from $x_j$ is specified by 2 parameters: 
$W_{ij}$ and $W_{ji}$.  When adding
a new dynamic variable, 
we first fix its parameters (to zero for $W$ and $\theta$
and one for $\tau$), and then allow them
to vary one at a time in the following order: 
$W_{ij},W_{ji},W_{ii},\tau_i,\theta_i$ 
(adding connections to every other $x_j$ one at a time).
An example is shown in Table~\ref{sigmoidTable}.

\begin{table}
\centering
\begin{tabular}{|c c l|}
\hline Model No.\ $i$ & Num. parameters $N_p$    
            &  ~Form of sigmoidal ODEs  \\ \hline
    0 & 3 & $\begin{aligned}[t]
        x_1(0) &= x_1^{\rm init} \\
        \frac{dx_1}{dt}&= -x_1/\tau_1 + W_{1 1} \xi(x_1) + W_{1 0} x_I
    \end{aligned}$ \\[30pt] \hline
    1 & 4 & $\begin{aligned}[t]
        x_1(0) &= x_1^{\rm init} \\
        \frac{dx_1}{dt}&= -x_1/\tau_1 + W_{1 1} \xi(x_1 + \theta_1) 
            + W_{1 0} x_I
    \end{aligned}$ \\[30pt] \hline
    2 & 6 & $\begin{aligned}[t]
        x_1(0) &= x_1^{\rm init} \\
        x_2(0) &= x_2^{\rm init} \\
        \frac{dx_1}{dt}&= -x_1/\tau_1 + W_{1 1} \xi(x_1 + \theta_1) 
            + W_{1 2} \xi(x_2) + W_{1 0} x_I \\
        \frac{dx_2}{dt}&= -x_2  
    \end{aligned}$ \\[76pt] \hline
    3 & 7 & $\begin{aligned}[t]
        x_1(0) &= x_1^{\rm  init} \\
        x_2(0) &= x_2^{\rm init} \\
        \frac{dx_1}{dt}&= -x_1/\tau_1 + W_{1 1} \xi(x_1 + \theta_1) 
            + W_{1 2} \xi(x_2) + W_{1 0} x_I \\
        \frac{dx_2}{dt}&= -x_2 + W_{2 0} x_I 
    \end{aligned}$ \\[76pt] \hline
    4 & 8 & $\begin{aligned}[t] 
        x_1(0) &= x_1^{\rm init} \\
        x_2(0) &= x_2^{\rm init} \\
        \frac{dx_1}{dt}&= -x_1/\tau_1 + W_{1 1} \xi(x_1 + \theta_1) 
            + W_{1 2} \xi(x_2) + W_{1 0} x_I \\
        \frac{dx_2}{dt}&= -x_2 + W_{2 1} \xi(x_1 + \theta_1) + W_{2 0} x_I
    \end{aligned}$ \\[76pt] \hline
    5 & 9 & $\begin{aligned}[t] 
        x_1(0) &= x_1^{\rm init} \\
        x_2(0) &= x_2^{\rm init} \\
        \frac{dx_1}{dt}&= -x_1/\tau_1 + W_{1 1} \xi(x_1 + \theta_1) 
            + W_{1 2} \xi(x_2) + W_{1 0} x_I \\
        \frac{dx_2}{dt}&= -x_2 + W_{2 2} \xi(x_2) 
            + W_{2 1} \xi(x_1 + \theta_1) + W_{2 0} x_I
    \end{aligned}$ \\[76pt] \hline
\end{tabular}
\caption{\label{sigmoidTable}
  The first six models of an 
  example model hierarchy in the sigmoidal class
  with one input $x_I$ and fixed $x_1^{\rm init}$ and $x_2^{\rm init}$.
}
\end{table}



For every adaptive fit model and the full multi-site phosphorylation
model,\footnote{ For the simple model fit to the phosphorylation data,
  parameters are always well-constrained and priors are unimportant,
  and we therefore do not use explicit priors.  } we use the same
prior for every parameter $\alpha_k$, which we choose as a normal
distribution ${\mathcal N}(0,10^2)$ with mean 0 and standard deviation
  $\varsigma = 10$.\footnote{ Some parameters ($\alpha$ and $\beta$ in
    the S-systems model class, $\tau$ in the sigmoidal model class,
    and $k$ and $K$ parameters in the full phosphorylation model) are
    restricted to be positive, which we accomplish by optimizing over
    the log of each parameter.  The priors are still applied in
    non-log space, effectively creating a prior that is zero for
    negative parameter values and $2 N(0,10)$ for positive parameter
    values.  }

\subsubsection{The law of gravity model}
\label{planetarySection}

For a mass $m$ in motion under the influence of the gravitational
field of a mass $M \gg m$, the distance $r$ between the two evolves as \cite{LL}
\begin{equation}
\frac{d^2 r}{d t^2} = \frac{h^2}{r^3} - \frac{G M}{r^2},
\end{equation}
where $h = (\vec v_0 \cdot \hat \theta) r_0$ is the specific angular
momentum, $\vec v_0$ is the initial velocity, $r_0$ is the initial
distance, $\hat \theta$ is the unit vector perpendicular to the line
connecting the two masses, and $G$ is the gravitational constant.
Setting the initial velocity parallel to $\hat \theta$ and measuring
distance in units of $\frac{GM}{v_0^2}$ and time in units of
$\frac{GM}{v_0^3}$, the dynamics become\footnote{
Note that $r_0$ sets the (conserved) angular momentum:
$h = \frac{GM}{v_0} r_0$ with $r_0$ in rescaled units.
}
\begin{equation}
\frac{d^2 r}{d t^2} = \frac{1}{r^2} \left( \frac{r_0^2}{r} - 1 \right).
\end{equation}

When written as two first-order differential equations, we see that
this system can be represented exactly in the S-systems class if the
particle does not fall onto the Sun:
\begin{eqnarray}
\nonumber
\frac{dr}{dt} &=& \chi - 1 \\
\frac{d \chi}{dt} &=& r_0^2 r^{-3} - r^{-2},
\end{eqnarray}
where we use the variable $\chi = \frac{dr}{dt} + 1$, so that the
resulting system's variables are never negative, a requirement of the
S-systems class.

\begin{figure}
\centering
\includegraphics[scale=1.25]{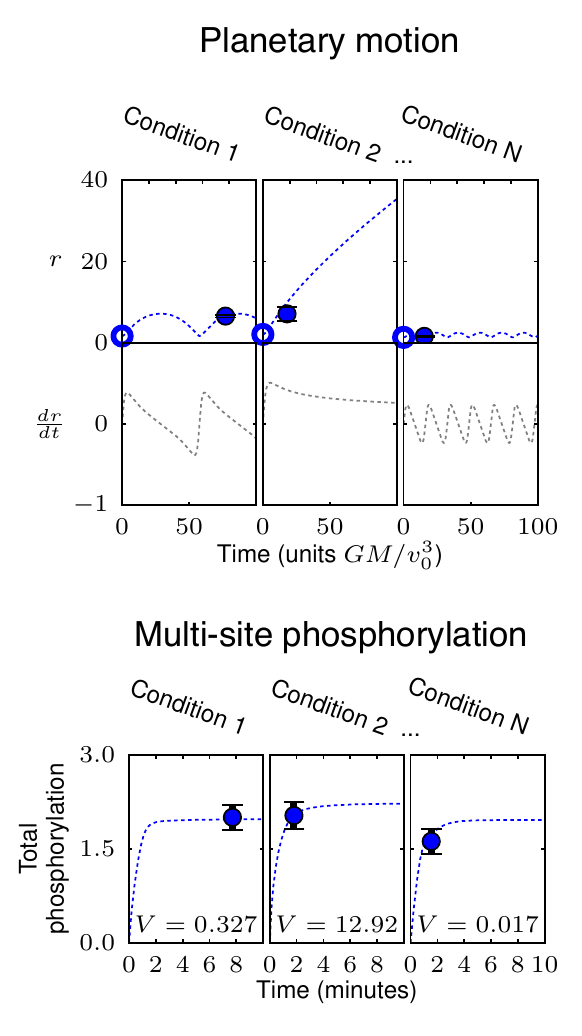}
\caption{\label{inSampleData}
  Typical in-sample data points for the planetary motion and
  multi-site phosphorylation model examples.  For the planetary
  motion, $r_0$ is treated as input, and for each in-sample $r_0$, $r$
  is measured, with added noise, at a single randomly chosen time
  between 0 and 100.  For multi-site phosphorylation, the single
  parameter $V$ is treated as input, and the total phosphorylation is
  measured, with added noise, at a single randomly chosen time between
  0 and 10 minutes.  Dotted lines show the original model behavior,
  filled circles with error bars show the in-sample data, and unfilled
  circles show the varying initial conditions in the planetary motion
  case.  The original planetary motion model includes a single hidden
  variable $X_2$ corresponding to the time derivative of $r$.  (For
  the yeast glycolysis example, a similar depiction of typical
  in-sample data is shown in the left panel of
  \figref{yeastModelSelection}.)  }
\end{figure}

To illustrate constructing an adaptive model for planetary motion, we
consider as input the initial distance from the sun $r_0$.  We sample
$r_0$ uniformly between 1 and 3 (in units of $GM/v_0^2$), which covers
the possible types of dynamics: at $r_0 = 1$, the orbit is circular;
when $1 < r_0 < 2$ the orbit is elliptical; when $r_0 = 2$ the orbit
is parabolic; and when $r_0 > 2$ the orbit is hyperbolic. In this and
later examples, to best determine the minimum number of measurements
needed for a given level of performance, we sample the system at a
single time point for each initial condition (\figref{inSampleData}),
rather than sampling a whole trajectory per condition.  This ensures
that samples are independent, which would not be the case for
subsequent data points of the same trajectory, and hence allows us to
estimate the data requirements of the algorithm more
reliably. Further, this is similar to the sampling procedure already
used in the literature \cite{SchValJen11}. In the planetary motion
case, we assume only the distance $r$ is measured, meaning the total
number of of datapoints $N_D = N$, where $N$ is the number of initial
conditions sampled.  We choose the time of the observation as a random
time uniformly chosen between $0$ and $100$, with time measured in
units of $GM/v_0^3$.  To each measurement we add Gaussian noise with
standard deviation equal to $5\%$ of the maximum value of $r$ between
$t = 0$ and $t = 100 ~GM/v_0^3$.

Typical training data for the model can be seen in
\figref{inSampleData}.  Fits to $N = 150$ data points are shown in
\figref{planetary}. Here our adaptive fitting algorithm selects a
model of the correct dimension, with one hidden variable.  The
selected model ODEs in this case are
\begin{eqnarray}
\nonumber
\frac{dr}{dt}   &=& e^{-3.405} r_0^{3.428}  r^{0.049}  X_2^{7.372}
                  - e^{-2.980} r_0^{2.936}  r^{0.046}  X_2{-4.925} \\
\frac{dX_2}{dt} &=&            r_0^{-0.651} r^{-3.435} X_2^{-0.014}
                  - e^{-0.006} r_0^{-4.288} r^{-1.595}.
\end{eqnarray}
Note that certain transformations of the hidden variable and
parameters can leave the output behavior unchanged while remaining in
the S-systems class.  First, the initial condition of hidden
parameters can be rescaled to 1 without loss of generality, so we
remove this degree of freedom and set $X_2(0) = 1$.  Second, we have
the freedom to let the hidden variable $X_2 \rightarrow X_2^\gamma$
for any $\gamma \neq 0$ with appropriate shifts in parameters.  To
more easily compare the fit model with the perfect model, in the
rightmost column of \figref{planetary} we plot $X_2^2$ on the vertical
axes instead of $X_2$ when comparing it to the dynamics of the true
hidden variable $\chi$.

\begin{figure}
\centering
\includegraphics[scale=1.25]{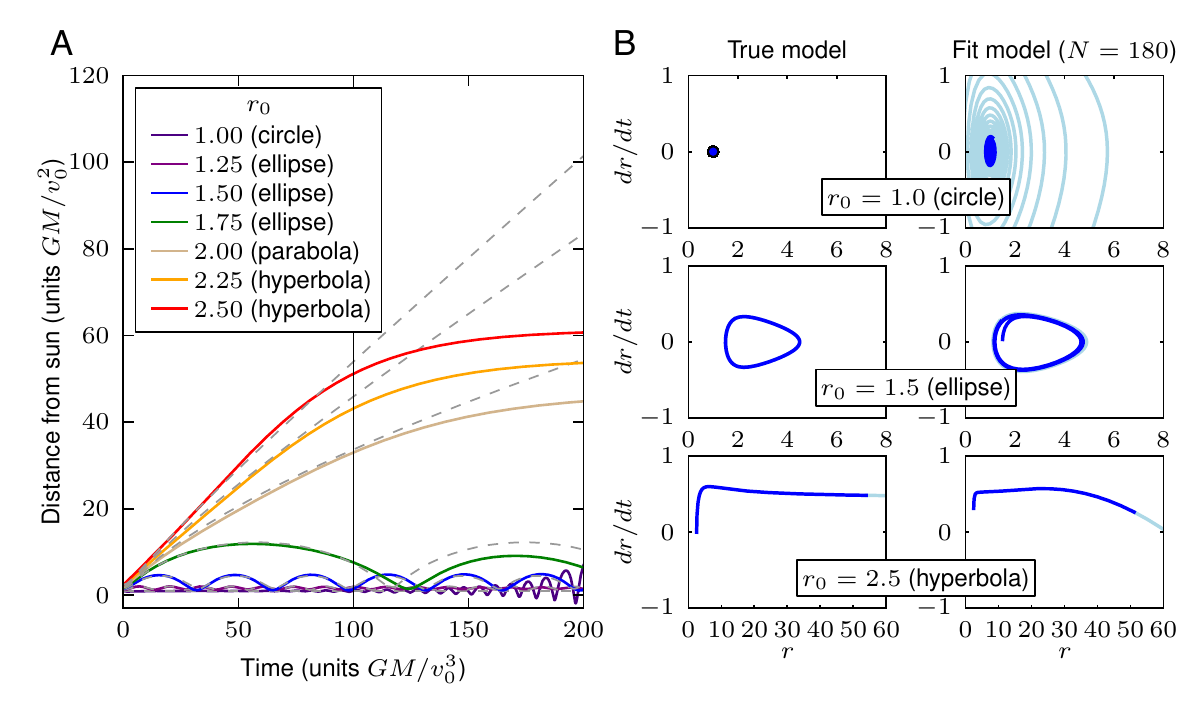}
\caption{\label{planetarySigmoidal}
  Fit of sigmoidal model to planetary data.  We know that the
  sigmoidal network model class is not likely to perform as well for
  the planetary data case because gravitational interactions do not
  saturate.  Here we show the performance of a model fit to $N = 180$
  data points, which contains three hidden variables.  The model still
  fits well in the time region where data is given (between 0 and 100
  $GM/v_0^3$, corresponding to the left half of A and the dark blue
  part of the trajectories in B), but has a larger divergence from the
  expected behavior at the extremes of the range of given $r_0$s in
  the extrapolated time region (corresponding to the right half of A
  and the light blue part of the trajectories in B). }
\end{figure}

Finally, we may compare performance when we fit the gravitation data
using sigmoidal models, a model class that we know is not
representative of the underlying mechanics.  The results are shown in
\figref{planetarySigmoidal}; the selected sigmoidal network,
which contains three hidden variables, still provides a
good fit to the data, as expected, but it does not generalize as well when
$r_0$ is near the edge of the range contained in the data and
timepoints are outside of the range of data to which they were
fit. This is expected since forces can diverge in the true law of
gravity, and they are necessarily limited in the sigmoidal
model.

\subsubsection{Multi-site phosphorylation model}

To explore a complicated biological system with relatively simple
output behavior, we imagine a situation in which an immune receptor
can be phosphorylated at each of five sites arranged in a linear
chain.  The rates of phosphorylation and dephosphorylation at each
site are affected by the phosphorylation states of its nearest
neighboring sites.  A site can be unphosphorylated ($U$) or
phosphorylated ($P$), and its state can change via one of two
processes.  The first process does not depend on states of neighboring
sites:
\begin{equation}
U_i \rightleftharpoons P_i,
\end{equation}
with on-rate $k_i^\mathrm{on}([U_i])$ and off-rate
$k_i^\mathrm{off}([P_i])$ that depend on the concentration of the
corresponding substrate.  The second, cooperative process 
happens only when a neighboring site $j$ is phosphorylated:
\begin{equation}
U_i P_j \rightleftharpoons P_i P_j
\end{equation}
with on- and off-rates $k_{ij}^\mathrm{on}([U_i P_j])$ and
$k_{ij}^\mathrm{off}([P_i P_j])$.  All rates $k$ are modeled as
Michaelis-Menten reactions: $k([S])~=~\frac{V [S]}{K_m + [S]}$.  With
each reaction specified by two parameters ($V$ and $K_m$) and 26
possible reactions, the phosphorylation model has a total of 52
parameters.  To more easily generate the differential equations that
govern the multi-site phosphorylation model, we use the BioNetGen
package~\cite{HlaFaeBli06,BioNetGen}.

When fitting this phosphorylation model, we use as input the parameter
$V_{23}^{\mathrm{on}}$, which is chosen from a uniform distribution in
log-space between $10^{-3}$ and $10^{3}$ min$^{-1}$.  The remaining 51
$V$ and $K_m$ parameters we sample randomly from our priors
on these parameters.  As output, we measure
the total phosphorylation of the 5 sites $P_{\mathrm{tot}}$ at a
single random time uniformly chosen between $0$ and $10$ minutes.  To
each measurement we add Gaussian noise with standard deviation equal
to $10\%$ of the $P_{\mathrm{tot}}$ value at $t = 10$ min.

Typical training data for the model is shown in
\figref{inSampleData}. The out-of-sample mean squared error, as
plotted in \figref{phosNdependence}, is measured over 100 new input
values selected from the same distribution as the in-sample values,
each of which is compared to the true model at 100 timepoints evenly
spaced from 0 to 10 minutes.

As a simple guess to the functional form of the total phosphorylation
timecourse as a function of our control parameter $V =
V_{23}^{\mathrm{on}}$ (the ``simple model'' in
\figref{phosNdependence}), we use an exponential saturation starting
at 0 and ending at a value $P_\infty$ that depends sigmoidally on $V$:
\begin{equation}
    P_{\mathrm{tot}} =
    P_\infty(V) \left[ 1 - \exp \left( \frac{t}{t_0} \right) \right],
\end{equation}
where
\begin{equation}
    P_\infty(V) =
    a + \frac{b}{2} \left[ 1 + \tanh \left( \frac{\log(V) - d}{c} \right) \right]
\end{equation}
and $a$, $b$, $c$, $d$, and $t_0$ are parameters fit to the
data. \figref{phosNdependence} shows that this simple {\em ad hoc}
model can fit the data quite well.

For the example shown in \figref{phosTimecourse}, the selected 
sigmoidal model consists of the ODEs
\begin{eqnarray}
\nonumber
\frac{dP_{\mathrm{tot}}}{dt} &=& \frac{-P_{\mathrm{tot}}}{e^{-1.219}} 
    + \frac{0.409}{1 + \exp(P_{\mathrm{tot}} - 4.469)}
    + \frac{7.087}{1 + \exp(X_2)} + 0.0005 V \\
\frac{dX_2}{dt} &=& -X_2 
    - \frac{2.303}{1 + \exp(P_{\mathrm{tot}} - 4.469)}
    - 0.071 V \\
\nonumber
X_2(0) &=& 0.101,
\end{eqnarray}
with $P_{\mathrm{tot}}(0) = 0$.

\begin{figure}
\centering
\includegraphics[scale=1.25]{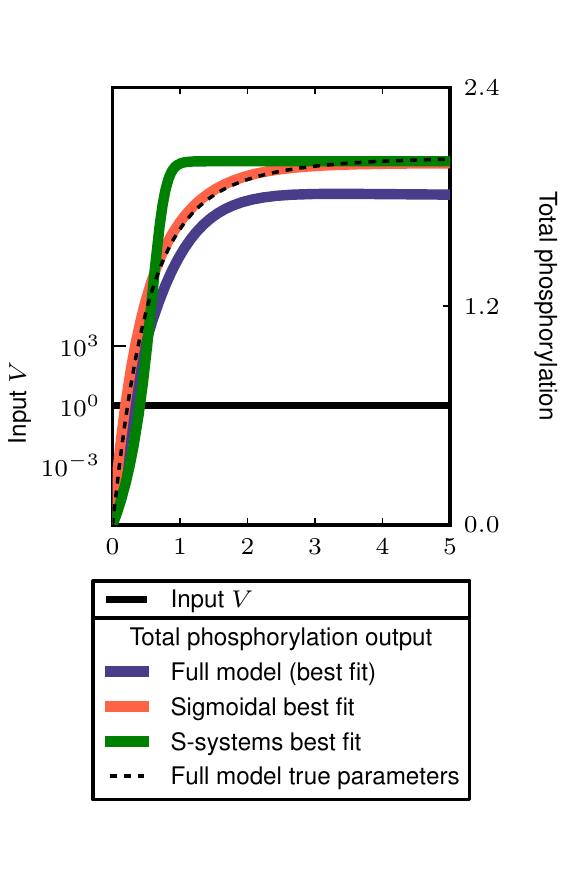}
\caption{\label{phosTimecourse_appendix}
A typical example of out-of-sample performance in the
multi-site phosphorylation example.  Here, each model is
fit using $N=50$ datapoints.  With this small amount of
data, the differences between model classes are more
apparent, with the sigmoidal model class clearly better 
predicting the dynamics 
than the S-systems model class and the full
phosphorylation model.
   }
\end{figure}

In this multi-site phosphorylation example, the sigmoidal model class
is a better performer than the S-systems class.  A typical example of
performance is depicted in \figref{phosTimecourse_appendix}.  
Though the S-systems
class makes predictions that are still qualitatively correct, and
its predictions steadily improve as $N$ increases, the
sigmoidal class comes closer to the true underlying model with an
equal amount of data.

The confidence intervals on the dynamics in \figref{phosTimecourse}
correspond to samples from the posterior over parameters given $N =
300$ data points.  In the notation of section \ref{derivationSection},
this posterior $P(\alpha ~|~ \mathrm{data}) \propto \exp{\left[-
    \tilde \chi^2(\alpha) / 2 \right]}$.  To generate samples from
this distribution, we use Metropolis Monte Carlo as implemented in
SloppyCell \cite{MyeGutSet07,SloppyCell}.  As a starting point, we use
the best-fit parameters from the model selection procedure, and we sample
candidate steps in parameter space from a 
multidimensional Gaussian corresponding to the
Hessian at the best-fit parameters.\footnote{ Unconstrained parameter
  directions in the proposal distribution, corresponding to singular
  values smaller than $\lambda_{\mathrm{cut}} =
  \lambda_{\mathrm{max}}/10$, where $\lambda_{\mathrm{max}}$ is the
  largest singular value, are cut off to $\lambda_{\mathrm{cut}}$ to
  produce reasonable acceptance ratios (near 0.5).  } From $10^4$
Monte Carlo steps, the first half are removed to avoid bias from the
initial condition, and every 50 of the remaining steps are used as 100
approximately independent samples from the parameter posterior.
We note that the median behavior over the Bayesian posterior is less
extreme than the behavior at the maximum likelihood parameters (not
shown), but still has fast-timescale dynamics indicative of
overfitting.  

\subsubsection{Yeast glycolysis model}

As an example of inference of more complicated dynamics, we use a
model of oscillations in yeast glycolysis, originally studied in terms
of temperature compensation \cite{RuoChrWol03} and since used as a
test system for automated inference \cite{SchValJen11}.  The model's
behavior is defined by ODEs describing the dynamics of the
concentrations of seven molecular species (the biological meaning of
the species is not important here):
\begin{eqnarray}
\frac{d S_1}{d t} &=& J_0 
    - \frac{k_1 S_1 S_6}{1 + (S_6/K_1)^q} \nonumber \\
\frac{d S_2}{d t} &=& 2 \frac{k_1 S_1 S_6}{1 + (S_6/K_1)^q}
    - k_2 S_2 (N-S_5) - k_6 S_2 S_5 \nonumber \\
\frac{d S_3}{d t} &=& k_2 S_2 (N-S_5) 
    - k_3 S_3 (A - S_6) \nonumber \\
\frac{d S_4}{d t} &=& k_3 S_3 (A - S_6) - k_4 S_4 S_5 
    - \kappa (S_4 - S_5) \label{yeastEqns} \\
\frac{d S_5}{d t} &=& k_2 S_2 (N-S_5) - k_4 S_4 S_5
    - k_6 S_2 S_5 \nonumber \\
\frac{d S_6}{d t} &=& -2 \frac{k_1 S_1 S_6}{1 + (S_6/K_1)^q} 
    + 2 k_3 S_3 (A - S_6) - k_5 S_6 \nonumber \\
\frac{d S_7}{d t} &=& \psi \kappa (S_4 - S_5) - k S_5. \nonumber
\end{eqnarray}
Parameter values, listed in Table~\ref{yeastParamsTable}, are set to
match with those used in Ref.~\cite{SchValJen11} and Table~1 of
Ref.~\cite{RuoChrWol03}, where our $S_5 = N_2$, our $S_6 = A_3$, and
our $S_7 = S_4^{ex}$.

\begin{table}
\centering
\begin{tabular}{ | c | l l |}
  \hline                        
  $J_0$   & 2.5    & mM min$^{-1}$           \\
  $k_1$   & 100.   & mM$^{-1}$ min$^{-1}$    \\
  $k_2$   & 6.     & mM$^{-1}$ min$^{-1}$    \\
  $k_3$   & 16.    & mM$^{-1}$ min$^{-1}$    \\
  $k_4$   & 100.   & mM$^{-1}$ min$^{-1}$    \\
  $k_5$   & 1.28   & min$^{-1}$              \\
  $k_6$   & 12.    & mM$^{-1}$ min$^{-1}$    \\
  $k$     & 1.8    & min$^{-1}$              \\
  $\kappa$& 13.    & min$^{-1}$              \\
  $q$     & 4      &                         \\
  $K_1$   & 0.52   & mM                      \\
  $\psi$  & 0.1    &                         \\
  $N$     & 1.     & mM                      \\
  $A$     & 4.     & mM                      \\
  \hline  
\end{tabular}
\caption{\label{yeastParamsTable}
  Parameters for the yeast glycolysis model 
  defined in Eqns.~(\ref{yeastEqns}).}
\end{table}

\begin{figure}
\centering
\includegraphics[scale=0.5]{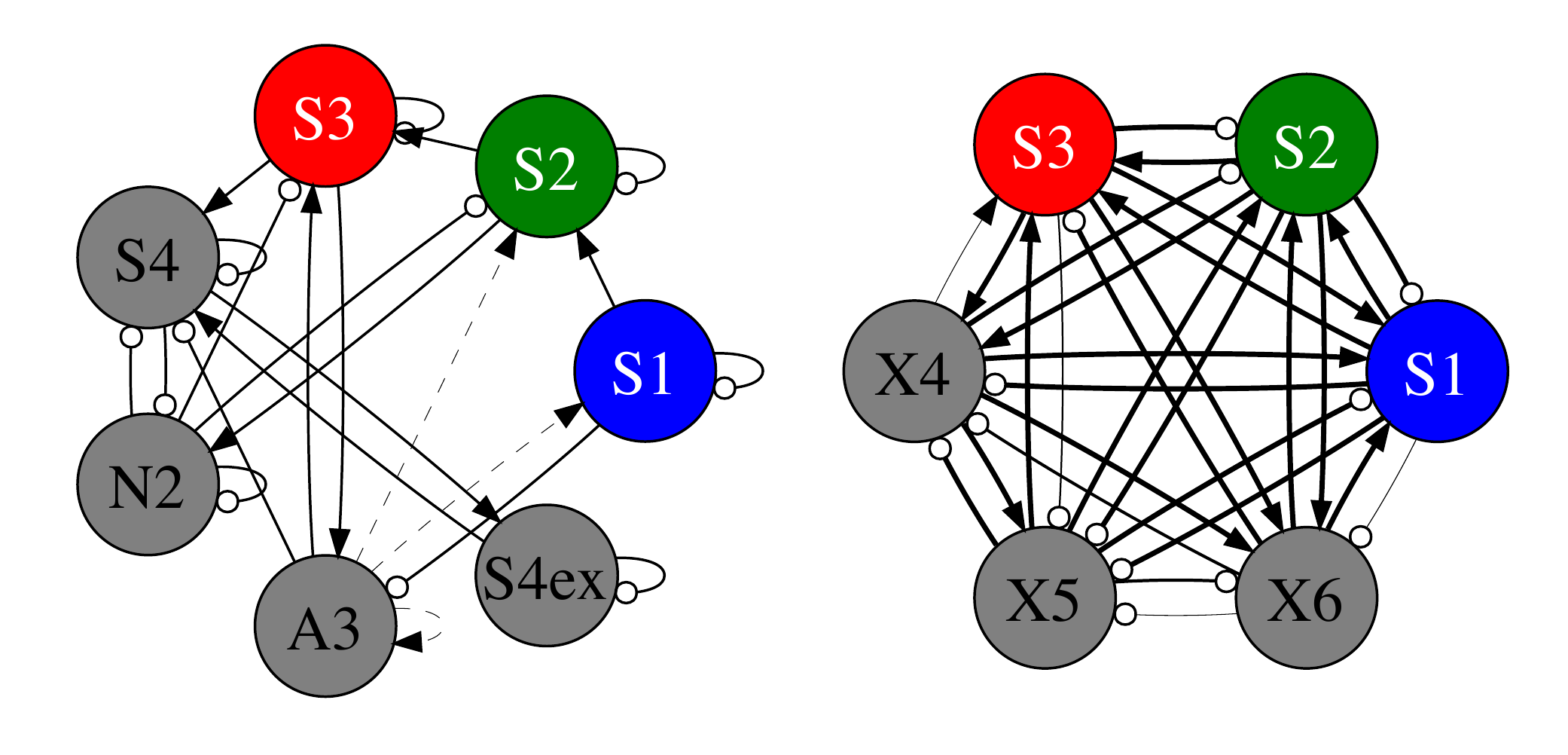}
\caption{\label{yeastModel}
  (Left) Network depicting the yeast glycolysis model defined by
  Eqns.~(\ref{yeastEqns}).  Solid arrows represent excitation, solid
  lines with circles represent inhibition, and dashed arrows represent
  other types of interaction terms.  (Right) Selected sigmoidal
  network fit to $N = 40$ noisy measurements from the yeast glycolysis
  model, as shown in \figref{yeastModelSelection}.  Again, arrows
  represent excitation and circles inhibition, with the thickness
  of arrows indicating interaction strength.  For clarity,
  self-inhibitory terms for each variable are not shown.  }
\end{figure}

\begin{table}
\centering
\begin{tabular}{ | c | l l l |}
  \hline 
  Variable  & In-sample IC (mM) & Out-of-sample IC (mM) & In-sample $\sigma$ (mM) \\
  \hline
  $S_1$     & [0.15, 1.60]      & [0.15, 3.05]          & 0.04872                 \\
  $S_2$     & [0.19, 2.16]      & [0.19, 4.13]          & 0.06263                 \\
  $S_3$     & [0.04, 0.20]      & [0.04, 0.36]          & 0.00503                 \\
  $S_4$     & 0.115             & 0.115                 & N/A                     \\
  $S_5$     & 0.077             & 0.077                 & N/A                     \\
  $S_6$     & 2.475             & 2.475                 & N/A                     \\
  $S_7$& 0.077             & 0.077                 & N/A                     \\
  \hline  
\end{tabular}
\caption{\label{yeastICTable}
  Initial conditions (IC) and standard deviations of 
  experimental noise ($\sigma$) 
  used in the yeast glycolysis model.
  Initial conditions for visible species $S_1$, $S_2$, and $S_3$
  are chosen uniformly from the given ranges, chosen to
  match Ref.~\cite{SchValJen11}.  Out-of-sample ranges are
  each twice as large as in-sample ranges.
  Initial conditions for the
  remaining hidden species are fixed at reference 
  initial conditions from 
  Refs.~\cite{SchValJen11} and \cite{RuoChrWol03}.
  In-sample noise is set at 10\% of the standard deviation of
  each variable's concentration in the limit cycle, as quoted in
  Ref.~\cite{SchValJen11}.}
\end{table}

For the yeast glycolysis model, we use as input the initial conditions
for the visible species $S_1$, $S_2$, and $S_3$.  These are each
chosen uniformly from ranges listed in the ``In-sample IC'' column of
Table~\ref{yeastICTable}.  Each of the three visible species are then
measured at a random time uniformly chosen from $0$ to $5$ minutes,
meaning the total number of datapoints $N_D = 3 N$ for this system,
where $N$ is the number of initial conditions sampled.  Gaussian noise
is added to each measurement with standard deviations given in
Table~\ref{yeastICTable}.  To evaluate the model's performance, we
test it using 100 new input values selected uniformly from the ranges
listed in the ``Out-of-sample IC'' column of Table~\ref{yeastICTable},
each of which is compared to the true model at 100 timepoints evenly
spaced from 0 to 5 min.  The correlation between the adaptive fit
model and the actual model over these 100 timepoints is calculated
separately for each visible species, set of initial conditions, and
in-sample data, and the average is plotted as the ``mean out-of-sample
correlation'' in \figref{yeastModelSelection}. The topology of the
selected network model is illustrated in \figref{yeastModel}.
Note that our model fitting approach assumes that
the model timecourse is fully determined
(aside from measurement error) by the
concentrations of measured species.
To be consistent with this assumption we do not vary
the initial conditions of the four hidden variables.
In future work it may be possible to relax this assumption,
allowing the current state of intrinsic variations in
hidden variables to be learned as well.


In Ref.~\cite{SchValJen11}, the EUREQa engine is used to infer
the same yeast glycolysis model that we use here.  We can roughly
compare performance as a function of computational and
experimental effort by measuring the number of required model
evaluations and measurements (\figref{yeastModelSelection}).
Here we compare the two approaches in more detail.

First, Ref.~\cite{SchValJen11} attempts to match time derivatives
of species concentrations as a function of species concentrations,
instead of species concentrations as a function of time as we do.
This means that each model evaluation\footnote{
    In our setup, we define a model evaluation as a single integration of
    the model ODEs (see Section~\ref{scalingSection}).
} is more computationally
costly for us, since it requires an integration of the ODEs over time.
It also means, however, that we are able to match well the phases of
oscillations, which remain unconstrained in Ref.~\cite{SchValJen11}.
The fitting of timecourses instead of derivatives also makes our
method focus on the fitting of dynamics near the
attractor, rather than attempting to constrain dynamics through
the entire phase space.

To consistently infer exact equations for the full 7-dimensional model,
Ref.~\cite{SchValJen11} used $20,000$ datapoints and 
roughly $10^{11}$ model evaluations.  We contrast this with our method that
produces reasonable inferred models using $40$ datapoints
and less than $5\times 10^8$ model evaluations (\figref{yeastModelSelection}).

\begin{figure}
\centering
\includegraphics[scale=1.]{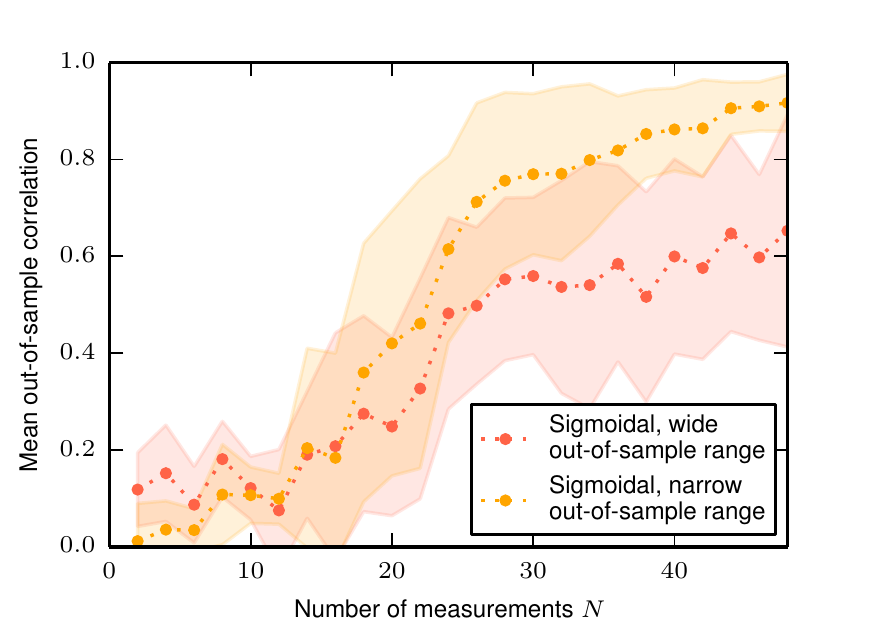}
\caption{\label{yeastVsN}
  Performance of inferred models of yeast glycolysis as a function of
  the number of measurements $N$.  Here we compare mean correlations
  produced for out-of-sample initial conditions chosen from ranges
  twice as large as in-sample ranges (``wide ranges,'' plotted in red)
  to when out-of-sample conditions are chosen from the same ranges as
  in-sample ranges (``narrow ranges,'' plotted in orange).  The mean
  and standard deviation over 5 realizations of in-sample data are shown
  by filled symbols and shaded regions.  }
\end{figure}

Finally, in the main text we test the performance of our yeast
glycolysis models for out-of-sample ranges of initial conditions that
are twice as large as the in-sample ranges from which data is taken,
as in Ref.~\cite{SchValJen11},
in order to more directly test their ability to extrapolate to regimes
that were not tested in training.  In \figref{yeastVsN}, we compare
this to performance when out-of-sample initial conditions are chosen
from the same ranges as in-sample data (note that, nonetheless, none
of the test examples has appeared in the training set).  Here we see
that the mean correlation can reach 0.9 using $N =
40$ measurements.

\subsection{Derivation of Bayesian log-likelihood estimate $\L$}
\label{derivationSection}

The derivation here largely follows Refs.~\cite{Bal97,BiaNemTis01}, but can
be traced to the 1970s \cite{Sch78}.  For a given model $M$ that depends
on parameters $\alpha$, our model selection algorithm requires an
estimate of the probability that $M$ is the model that produced a
given set of data $\set{y_i}$ with corresponding error estimates
$\set{\sigma_i}$ (measured at a set of timepoints $\set{t_i}$), and
$i=1,\dots,N$, so that there are $N$ measurements.  Since the
parameters $\alpha$ are unknown aside from a prior distribution
$P(\alpha)$, we must integrate over all possible values:
\begin{align}
P( M ~|~ \mathrm{data} ) 
    &= P( M ~|~ \set{y_i,\sigma_i,t_i} ) \\
    &= Z^{-1}_\alpha \int d^{N_p} \alpha ~
            P( M ~|~ \set{y_i,\sigma_i,t_i};\alpha ) ~ P(\alpha),
\end{align}
where the normalization constant 
$Z_\alpha = \int d^{N_p} \alpha ~ P(\alpha)$ and $N_p$ is the
number of parameters.
In terms of the output given the model, Bayes rule states
\begin{equation}
P\left( M ~|~ \set{y_i,\sigma_i,t_i};\alpha \right) = 
    \frac{ P\left( M \right) } { P\left(\set{y_i}\right) }
        P\left( \set{y_i} ~|~ M(\alpha);\set{\sigma_i,t_i} \right).
\end{equation}
Assuming that the model output has normally distributed measurement errors,
\begin{align}
P\left( \set{y_i} ~|~ M(\alpha);\set{\sigma_i,t_i} \right) &= 
    \prod_{i=1}^{N} P\left( y_i ~|~ M(\alpha);\sigma_i;t_i \right) \\ \nonumber
    &= \prod_{i=1}^{N} \frac{1}{\sqrt{2\pi \sigma_i^2}} 
        \exp{\left[ -\frac{1}{2} 
        \left( \frac{y_i - M(t_i,\alpha)}{\sigma_i} \right)^2 \right]} \\ \nonumber
    &= Z^{-1}_\sigma   
      \exp{\left[  -\frac{1}{2} \sum_{i=1}^{N}
        \left( \frac{y_i - M(t_i,\alpha)}{\sigma_i} \right)^2 \right]} \\ \nonumber
    &= Z^{-1}_\sigma 
      \exp{ \left[ -\frac{1}{2} \chi^2(M(\alpha),\set{y_i,\sigma_i,t_i}) \right] },
\end{align}
where $\chi^2$ is the usual goodness-of-fit measure consisting of the sum 
of squared residuals, and $Z_\sigma$ is the normalization constant
$\prod_{i=1}^{N} \sqrt{2\pi \sigma_i^2}$.
Thus we have:\footnote{
    We simplify notation by letting 
    $\chi^2(\alpha) = \chi^2(M(\alpha),\set{y_i,\sigma_i,t_i})$.
}
\begin{equation}
P( M ~|~ \mathrm{data} ) = 
    C Z^{-1}_\alpha \int d^{N_p} \alpha ~
    ~ \exp{ \left[ -\frac{1}{2} \tilde \chi^2(\alpha) \right] },
\end{equation}
where $C \equiv 2 P(M) / Z_\sigma P\left(\set{y_i}\right)$ and $\tilde
\chi^2(\alpha) = \chi^2(\alpha) - 2 \log P(\alpha)$.  Since we will be
comparing models fitting the same data, and we assume all models have
the same prior probability $P(M)$, $C$ will be assumed constant in all
further comparisons (but see Ref.~\cite{WolMac97} for the discussion
of this assumption).

If there are enough data to sufficiently constrain the parameters (as
is the case for ideal data in the limit $N \rightarrow \infty$), then
the integral will be dominated by the parameters near the single set
of best-fit parameters $\alpha_\best$.  To lowest order in $1/N$, we
can approximate the integral using a saddle-point approximation
\cite{BiaNemTis01}:
\begin{align}
P( M ~|~ \mathrm{data} )
    \label{Pintegral}
    &\approx C Z^{-1}_\alpha
    \exp{\left[-\frac{1}{2} \tilde \chi^2(\alpha_\best)  \right]}
    \int d^{N_p} \alpha ~ \exp{ \left[ -
        (\alpha - \alpha_\best) \H (\alpha - \alpha_\best) \right]},
    \end{align}
where $\H$ is the Hessian:\footnote{
    Near the best-fit parameters where residuals are small, and
    when priors are Gaussian,
    $\H$ is approximated by the Fisher Information Matrix,
    which depends only on first derivatives of model behavior:
    $\H \approx J^TJ + \Sigma^{-2}$, where the Jacobian  
    $J_{i \ell} = \frac{1}{\sigma_i} \frac{\partial M_i}{\partial \alpha_{\ell}}$ and
    the diagonal matrix 
    $\Sigma^{-2}_{k \ell} = \delta_{k \ell} \varsigma_k^{-2}$ 
    expresses the effects of parameter priors.
}
\begin{equation}
\label{Hessian}
\H_{k\ell} = \frac{1}{2} 
    \frac{\partial^2 \tilde \chi^2(\alpha)}{\partial \alpha_k d \alpha_\ell}
    \bigg|_{\alpha_\best}.
\end{equation}
If we assume normally distributed priors on parameters
with variances $\varsigma^2_k$,
the log posterior probability becomes
\begin{equation}
\log P( M ~|~ \mathrm{data} ) \approx
    \mathrm{const} - \frac{1}{2} \tilde \chi^2(\alpha_\best)
    - \frac{1}{2} \sum_{\mu=1}^{N_p} \log \lambda_\mu
    - \frac{1}{2} \sum_{k=1}^{N_p} \log \varsigma^2_k,
\end{equation}
where $\lambda_\mu$ are the eigenvalues of $\H$, and the last term
comes from $Z_\alpha$.  We thus use as our measure of model quality
\begin{equation}
\label{L}
\L \equiv - \frac{1}{2} \tilde \chi^2(\alpha_\best)
    - \frac{1}{2} \sum_{\mu} \log \lambda_\mu
    - \frac{1}{2} \sum_k \log \varsigma^2_k.
\end{equation}
Eq.~\eqref{L} is a generalization of the Bayesian Information
Criterion (BIC) \cite{Sch78} when parameter sensitivities and priors
are explicitly included.\footnote{For well-constrained parameters, we
  expect, to lowest order in $1/N$, our result to be equal to the BIC
  result of $-\frac{1}{2} \tilde \chi^2(\alpha_\best) + \frac{1}{2}
  N_p \log{N}$.  } The first term is the familiar $\chi^2$ ``goodness
of fit,'' and the last two terms constitute the fluctuation
``penalty'' for overfitting or complexity. Note that here the goodness
of fit and the complexity penalty are both functions of the entire
dynamics, rather than individual samples, which is not a common
application of Bayesian model selection techniques.

\subsection{Fitting algorithm}

We are given $N$ data points $\vecx_i$ at known times $t_i$ and known
exogenous parameters $I_i$, and with known or estimated variances
$\sigma_i^2$.  We are approximating the functions $\vec{F}_X$ and
$\vec{F}_Y$ in Eq.~(\ref{eq:model}),  where $\vecy$ are hidden dynamic
model variables, and $\vecx = \vecx(t,I)$ and $\vecy = \vecy(t,I)$ in
general depend on time $t$ and inputs $I$.  As described in
Section~\ref{derivationSection}, we fit to the data $\vecx_i$ using a
combination of squared residuals from the data and priors $P(\alpha)$
on parameters $\alpha$, which we assume to be Gaussian and centered at
zero:
\begin{equation}
\label{noDerivChiSq}
\tilde \chi^2 = \sum_{i=1}^N
    \left( \frac{\vecx_i - \vecx(t_i,I_i)}{\sigma_i} \right)^2
    + 2 \sum_{k=1}^{N_p} \left( \frac{\alpha_k}{\varsigma_k} \right)^2,
\end{equation}
where $F$'s are integrated to produce the model values $\vecx$ and
$\vecy$:
\begin{eqnarray}
  \vecx(t,I) &=& \vecx_0(I) + \int_0^t \vec{F}_X(\vecx(s,I),\vecy(s,I))~ds \\
  \vecy(t,I) &=& \vecy_0(I) + \int_0^t \vec{F}_Y(\vecx(s,I),\vecy(s,I))~ds.
\end{eqnarray}

To fit parameters, we use a two step process akin to simulated
annealing that uses samples from a ``high temperature'' Monte Carlo
ensemble as the starting points for local optimization performed using
a Levenberg-Marquardt routine.  The phenomenological models are
implemented using SloppyCell \cite{MyeGutSet07,SloppyCell} in order to
make use of its parameter estimation and sampling routines.

Following is a high-level description of the fitting algorithm, with
choices of parameters for the examples in the main text listed in
Table~\ref{hyperParamsTable}.

\begin{enumerate}
\item Choose a model class, consisting of a sequence of nested models
  indexed by $i$, where the number of parameters $N_p$ monotonically
  increases with $i$. Choose a step size $\Delta p$.
\item Given data at $N_\mathrm{total}$ timepoints, fit to data from
  the first $N$ timepoints, where $N$ is increased to
  $N_\mathrm{total}$ in steps of $\Delta N$.
\item At each $N$, test models of increasing number of parameters
  $N_p$ (stepping by $\Delta p$) until $\L$ consistently decreases
  (stopping when the last $i_\mathrm{overshoot}$ models tested have
  smaller $\L$ than the maximum).  For each model, to calculate $\L$:
    \begin{enumerate}
    \item Generate an ensemble of starting points in parameter space
      using Metropolis-Hast\-ings Monte Carlo to sample from
      $P(\alpha) \propto \exp(-\tilde \chi^2(\alpha)/ 2 T N_D)$ with
      $\tilde \chi^2$ from \eqref{noDerivChiSq}.  The temperature $T$
      is set large to encourage exploration of large regions of
      parameter space, but if set too large can result in a small
      acceptance ratio.  Infinities and other integration errors are
      treated as $\tilde \chi^2 = \infty$.
        \begin{enumerate}
        \item Use as a starting point
        the best-fit parameters from a smaller $N_p$
        if a smaller model has been previously fit, 
        or else default parameters.  
        \item As a proposal distribution for candidate steps
        in parameter space, use an
        isotropic Gaussian with standard deviation 
        $\sqrt{T N_D}/\lambda_{\mathrm{max}}$, 
        where $N_D$ is the total number of data residuals and
        $\lambda_{\mathrm{max}}$
        is the largest singular value of the Hessian
        [Eq.~\eqref{Hessian}] at the starting parameters.
        \item If this model has 
        previously been fit to less data, use those 
        parameters as an additional member of the ensemble.
        \end{enumerate}
      \item Starting from each member of the ensemble, perform a local
        parameter fit, using Levenberg-Marquardt to minimize $\tilde
        \chi^2$ from \eqref{noDerivChiSq}.  Stop when convergence is
        detected (when the L1 norm of the gradient per parameter is
        less than {\tt avegtol}) or when the number of minimization
        steps reaches {\tt maxiter}.  The best-fit parameters
        $\alpha^*$ are taken from the member of the ensemble with the
        smallest resulting fitted $\tilde \chi^2$.
    \item At $\alpha^*$, calculate $\L$ from \eqref{L}.
    \end{enumerate}
  \item For each $N$, the model with largest log-likelihood $\L$ is
    selected as the best-fit model.
\end{enumerate}

\begin{table}
\centering
\begin{tabular}{| l | l |}
  \hline
  $\Delta p$ (gravitation and phosphorylation examples)   & $2$       \\
  $\Delta p$ (yeast example)                            & $5$       \\
  $i_\mathrm{overshoot}$                                & $3$       \\
  \hline
  Ensemble temperature $T$ (full phosphorylation model)\footnotemark[1]
  & $10$          \\
  Ensemble temperature $T$ (all other models) & $10^3$              \\
  Total number of Monte Carlo steps (full phosphorylation model)\footnotemark[1]
  & $10^2$        \\
  Total number of Monte Carlo steps (all other models)  & $10^4$    \\
  Number of ensemble members used                       &  $10$     \\
  \hline
  {\tt avegtol}                                         & $10^{-2}$  \\
  {\tt maxiter}                                         & $10^2$     \\
  \hline
\end{tabular}
\caption{\label{hyperParamsTable} Adaptive inference algorithm
    parameters. $^1$In the full phosphorylation model,
    we fit parameters in log-space since they are known to be 
    positive.  This makes the model more
    sensitive to large changes in parameters, meaning that we are forced
    to be more conservative with taking large steps in parameter
    space to achieve reasonable acceptance ratios. 
  }
\end{table}


\subsection{Scaling of computational effort}
\label{scalingSection}

\begin{figure}
\centering
\includegraphics[scale=1.0]{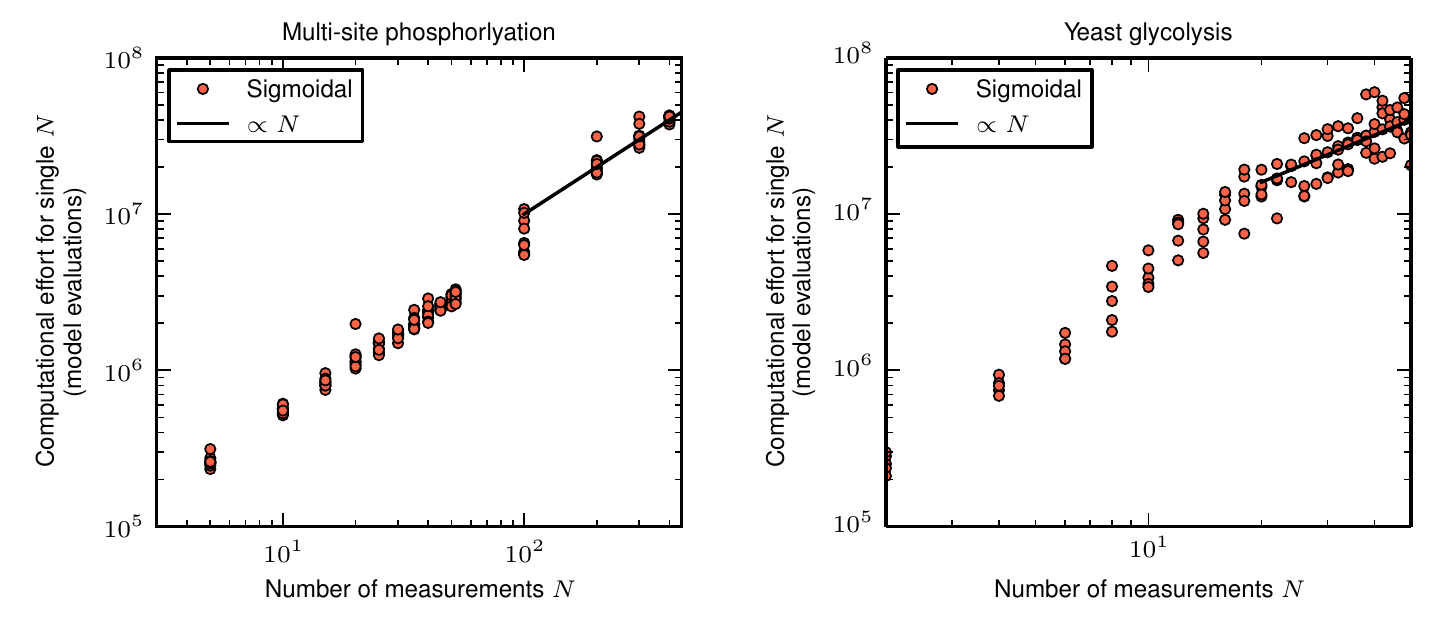}
\caption{\label{modelEvaluations}
  The number of model evaluations (integrations) used at each $N$, for
  the multi-site phosphorylation and yeast glycolysis examples.  Once
  the size of model has saturated, we expect the number of evaluations
  to scale linearly with $N$ (black lines).  If the selected model size is
  growing with $N$, as in the yeast glycolysis example below $N=20$
  (see \figref{yeastParameters}), we expect faster than linear growth.
}
\end{figure}

\begin{figure}
\centering
\includegraphics[scale=1]{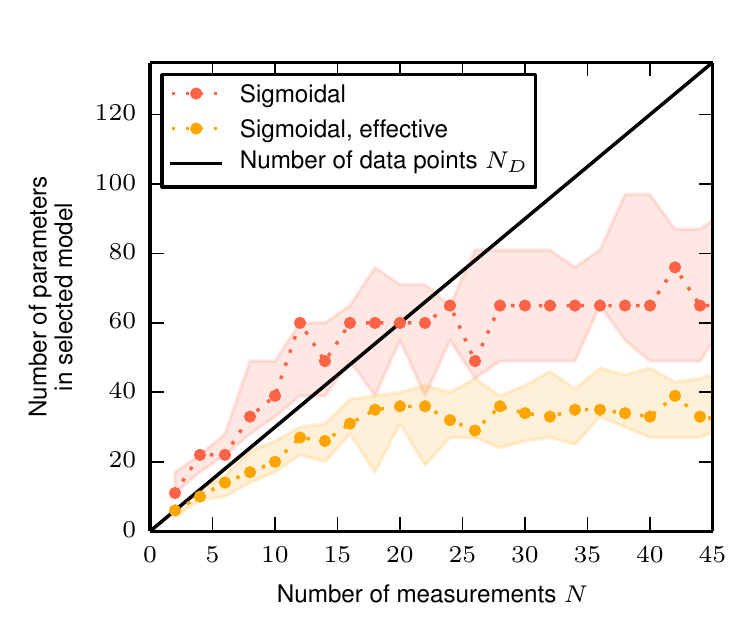}
\caption{\label{yeastParameters}
Fitting sigmoidal models to the 
yeast glycolysis oscillation data, the
number of total parameters in the selected model, 
plotted in red,
saturates to roughly 65.  The solid line compares
the number of parameters in the selected model
to the number of data points $N_D$ used to infer the model.
In orange, we plot the effective number of 
parameters, which we define as the number of directions in parameter
space that are constrained by the data such
that the corresponding Hessian eigenvalue
$\lambda > 1$ (compared to parameter priors with
eigenvalue $10^{-2}$).  We expect the optimal
effective number of parameters to stay
below $N_D$.  Shown are the median and full range of
values over 5 data realizations.
}
\end{figure}

In \figref{modelEvaluations}, we plot the number of model evaluations
used in each search for the best-fit phenomenological model.  We
define a model evaluation as a single integration of a system of
ODEs.\footnote{ Note that the amount of necessary CPU time per
  integration is dependent on the size and stiffness of the system.  }
This includes both integration of model ODEs and the derivatives of
model ODEs, used in gradient calculations.\footnote{ The number of
  integrations per gradient calculation is proportional to the number
  of parameters.  This means that the computational effort used to fit
  large models is dominated by gradient calculations.  } Note that in
\figref{yeastModelSelection}, to indicate the total number of
evaluations used as $N$ is gradually increased to its final value, we
plot the cumulative sum of the number of model evaluations depicted in
\figref{modelEvaluations}. We see that the number of model evaluations
scales superlinearly with $N$ if the selected model size is growing
with $N$, as is the case in the yeast glycolysis model below about $N
= 20$ (\figref{yeastParameters}).  When the model size saturates, the
number of model evaluations scales roughly linearly with $N$.

\putbib[PredictivePhenomenology]

\end{bibunit}



\end{document}